\documentclass[showpacs,amsmath,amssymb,aps,pra,preprint]{revtex4-1}

\usepackage{xcolor}
\usepackage{times}
\usepackage{graphicx}
\usepackage[english]{babel}
\usepackage{hyperref}

\usepackage{amsfonts,amssymb,amsmath,braket}

\usepackage[latin1]{inputenc}

\begin{document}
\title{Controlling the Velocities and Number of Emitted Particles in the Tunneling to Open Space Dynamics}

\author{Axel U. J. Lode$^{1,\ast}$\footnote[0]{$^\ast$ Corresponding author}, Shachar Klaiman$^{1}$,  Ofir. E. Alon$^2$,\\
Alexej I. Streltsov$^1$, and Lorenz S. Cederbaum$^1$}
\affiliation{$^1$ Theoretische Chemie,
\normalsize{Physikalisch-Chemisches Institut,}
\normalsize{Universit\"at Heidelberg,}
\normalsize{Im Neuenheimer Feld 229, D-69120 Heidelberg}}
\affiliation{
\normalsize{$^2$ Department of Physics,}
\normalsize{University of Haifa at Oranim,}
\normalsize{Tivon 36006, Israel}}

\begin{abstract}
A scheme to control the many-boson tunneling process to open space is derived and demonstrated. The number of ejected particles and their velocities can be controlled by two parameters, the threshold of the potential and the interparticle interaction. Since these parameters are fully under experimental control, this is also the case for the number of ejected particles and their emission spectrum. The process of tunneling to open space can hence be used, for example, for the quantum simulation of complicated tunneling ionization processes and atom lasers. To understand the many-body tunneling process, a generalization of the model introduced in [Proc. Natl. Acad. Sci. USA, {\bf 109}, 13521 (2012)] for tunneling in the absence of a threshold is put forward and proven to apply for systems with a non-zero threshold value. It is demonstrated that the model is applicable for general interparticle interaction strengths, particle numbers and threshold values. The model constructs the many-body process from single-particle emission processes. The rates and emission momenta of the single-particle processes are determined by the chemical potentials and energy differences to the threshold value of the potential for systems with different particle numbers. The chemical potentials and these energy differences depend on the interparticle interaction. Both the number of confined particles and their rate of emission thus allow for a control by the manipulation of the interparticle interaction and the threshold. Numerically exact results for two, three and one hundred bosons are shown and discussed. The devised control scheme for the many-body tunneling process performs very well for the dynamics of the momentum density, the correlations, the coherence and of the final state, i.e., the number of particles that remain confined in the potential.
\end{abstract}


\maketitle

\clearpage
\section{Introduction}
The understanding of quantum many-body dynamics has been pushed forward in recent years by the realization of and unique possibilities to control Bose-Einstein condensates (BECs) in the laboratory \cite{oberthaler,greiner:02,BECketterle,anderson:95,bradley:95}. The confinement of BECs \cite{Henderson:09}, their interparticle interactions \cite{chin:10}, 
and their dimensionality \cite{1d1,1d2,1d3} can be manipulated in experiments almost at will. Using these extensive mechanisms of control, 
BECs are used as so-called quantum simulators to study a variety of physical systems: Solid state systems are studied with optical lattices \cite{bloch:08,oberthaler,OL_solid_state1,OL_solid_state2} and even problems in astrophysics are tackled \cite{steinhauer:10,macher:09,westbrook:12}.

One of the fundamental phenomena of quantum mechanics is the tunneling process. Despite lacking the energy to overcome a potential barrier, quantum particles are able to escape by \textit{tunneling} through the barrier. This is due to the probabilistic nature of quantum mechanics: The particles have a non-zero probability to be found on the other side of the barrier. The physics of tunneling for a single particle is well-understood \cite{tun_book} and were described already in the 20s of the previous century, see Refs.~\cite{WKB,tunstart1,tunstart2}. The corresponding many-body process of interacting particles has also been studied, see, e.g., Refs.~\cite{Brand:11,LINCOLN,TGtunneling,muga2,PhysRevA.87.043626,MQTBECWKB,lenztunfrag,nonexpCS,axel:09,*axel:10,axel:12}.
In the case of many-body tunneling, the mechanism of the dynamics has only been revealed recently \cite{axel:12}. The many-body tunneling process to open space is built up from many simultaneous single-particle emission processes. The velocities of the emitted bosons are defined by the chemical potentials of trapped interacting systems of different particle number with large accuracy.

The key many-body features of the tunneling process are the gradual loss of initial coherence which manifests in the occurrence of fragmentation -- an abundant phenomenon in the eigenstates and dynamics of many-boson systems, see, e.g., Refs.~\cite{nozieres:82,nozieres:96,spekkens:99,alon.prl2:05,lenzexact2,fragmentothers2,streltsov:06}. Explicitly, the emitted particles lose the coherence with both the trapped source and among each other. In order to monitor and derive the mechanism  by which the coherence is lost while tunneling to open space, normalized correlation functions as introduced by Glauber in Refs.~\cite{Glauber,Glauber1} prove to be the best quantities of analysis. 

The study of the many-body physics of tunneling to open space in Ref.~\cite{axel:12} used a novel quantum many-boson method, the multiconfigurational time-dependent Hartree method for bosons (MCTDHB) \cite{streltsov:06,alexejsplit}. MCTDHB provides the means to solve the time-dependent many-boson Schr\"odinger equation numerically exact for a wide range of problems, see, e.g., \cite{axel_exact,Chaos:12,axel:12,sakmann:09,sakmann.pra:10}.

In the present work it is shown that the many-boson tunneling to open space dynamics can be controlled extensively by a slight manipulation to the setup described in the study that revealed the mechanism of the ongoing tunneling dynamics \cite{axel:12}, namely the addition of a potential threshold. The physics of the single-particle processes which assemble the many-body tunneling dynamics are determined entirely by the chemical potentials of trapped subsystems with different particle number and the available kinetic energy after the emission. Hence the process can be controlled by manipulating the threshold $T$ which affects the available kinetic energy after emission and the interaction $\lambda_0$ which affects the chemical potentials.

How one can change and control the tunneling process of a many-body system as prescribed in \cite{axel:12}? In the present case this is achieved by the following two measures: Introducing a threshold to the one-body potential of the Hamiltonian of the system, i.e., setting it to a constant value $T$ in the asymptotic region, and altering the interparticle interaction strength $\lambda_0$. By altering the threshold, the momenta of the emitted particles can be managed and bound states may be created. Such a creation of a bound state of a finite number of particles implies that the final state of the process has been altered by adjusting $T$. In this altered final state, a controlled number of particles stays confined in the reservoir and the remainder of the many-body system escapes to open space. Furthermore, by altering the interparticle interaction strength $\lambda_0$, the chemical potentials driving the simultaneous single-particle processes (cf. Fig.
~\ref{pot-tnot0}) 
can be altered and their rates controlled.

In the wider context of quantum simulators the system is related to atom lasers and ionization processes: the modification of the threshold $T$ allows an investigation of different ionization thresholds and the interaction strength $\lambda_0$ can be used to tune characteristic velocities of the emission and distances of the peaks in the momentum distribution. The free part of the potential resembles the situation in atom laser experiments \cite{BECketterle,atoml1,atoml2,atoml3}.  
This extensive control described above might allow one to study the coherence dynamics of atom lasers \cite{BECketterle,atoml1,atoml2,atoml3} and potentially complicated tunneling ionization processes \cite{photoass_tun,distun} which are not amenable for in-detail experiments. This is because the momentum distributions in the many-boson tunneling process can be tuned almost at will. Hence, the many-body tunneling process of ultracold bosonic atoms could be used as a quantum simulator for a broad range of these processes.

The protocol for the process is as follows: The ground state of an interacting system in a parabolic trap is prepared, then the potential is transformed abruptly to an open shape and finally the dynamics are analyzed from a many-body perspective for different interaction strengths $\lambda_0$ and particle numbers $N$ (cf. Refs.~\cite{axel:09,*axel:10,axel:12}). The final form of the potential now has a nonzero asymptotic value $T$. To properly describe and assess the impact of this threshold on the occurring dynamics it is instructive, to first find a suitable smooth shape for the potential (Section~\ref{setup}) and then to analyze the energetics in the new potential (Section~\ref{modelv2}). This is done starting from a conjecture for the emission momenta, relevant energies and chemical potential described in Ref.~\cite{axel:12} and the model of the many-body process described therein. The strategy pursued here is to start from the simplest case of $N=2$ bosons (Section~\ref{cont2}) using only the threshold 
$T$ as a control parameter. Henceforth, the control possibilities for the dynamics of $N=3$ bosons with the interparticle interaction strength $\lambda_0$ are explored (Section~\ref{cont3}). The relation of the control parameters ($\lambda_0$ and $T$) to the available final states in the problem is subsequently used to control the emission momenta, chemical potentials, and the number of emitted particles for a many-boson 
system composed of $N=101$ bosons (Section~\ref{contmany}).
It turns out, that the control parameters, i.e., the threshold $T$ and the interaction strength $\lambda_0$, are sufficient to exert a big amount of control on the final state and even the correlation dynamics of the many-body process. Summary and outlook are found in Section~\ref{fin}.

\section{Hamiltonian and quantities of analysis}\label{setup}
The setup of the system is depicted in Figs.~\ref{pot-tnot0} and \ref{model_mu_T}. It is similar to the setup of the tunneling with zero threshold as described in Refs.~\cite{axel:12,axel:09,axel:10}. In this section the Hamiltonian and potential are introduced and an outline of the changes with respect to the potential used for the tunneling dynamics with zero threshold is given. Furthermore, the quantities by which the many-body dynamics will be analyzed throughout the present study are given.

\subsection{The Hamiltonian}
The tunneling process of ultracold many-body systems to open space is described by the time-dependent Schr\"odinger equation,
\begin{equation}
 i\partial_t \vert \Psi \rangle = \hat{H} \vert \Psi \rangle. \label{TDSE}
\end{equation}
Here, $\vert \Psi \rangle$ is the wave function which depends on the spatial coordinates of all particles and $\hat{H}$ is the many-body Hamiltonian,
\begin{equation}
 \hat{H}=\sum_{i=1}^N \hat{h}_i + \sum_{i<j=1}^N \hat{W}_{ij}. \label{HAM}
\end{equation}
For ultracold atomic bosons $\hat{H}$ contains one-body operators $\hat{h}_i$ for each boson and two-body operators $\hat{W}_{ij}$ for every pair of particles. 
The key to MCTDHB's efficiency in solving Eq.~\eqref{TDSE} exactly numerically \cite{axel_exact} lies in the usage of a time-dependent, variationally optimized many-body basis set. The introduction of the method and computational details are deferred to Appendix~\ref{MCTDHB} for brevity.

It remains to specify the considered Hamiltonian. 
For convenience, dimensionless units are used. This means that $\hat{H}$ is devided by $\frac{\hbar^2}{L^2 m}$, where $\hbar$ is Planck's constant, $m$ is the mass of the considered particles and $L$ is a length scale that one introduces. The two-body operators $\hat{W}_{ij}$ in the Hamiltonian \eqref{HAM} of ultracold bosonic systems mitigate the particle-particle interaction. Since the interaction is well described by s-wave scattering, a contact interaction potential, $\hat{W}_{ij}= \lambda_0 \delta(x_j-x_i)$ is an appropriate description. The parameter $\lambda_0=\frac{2m\omega_{\perp}L}{\hbar} a_s$ is related to the s-wave scattering length $a_s$ in the ultracold atomic sample. It can be tuned with the aid of Feshbach resonances \cite{chin:10} or the trap geometry, i.e., the transversal confinement frequency $\omega_\perp$.

\begin{figure}[!]
 \begin{center}
  \includegraphics[angle=-90,width=\textwidth]{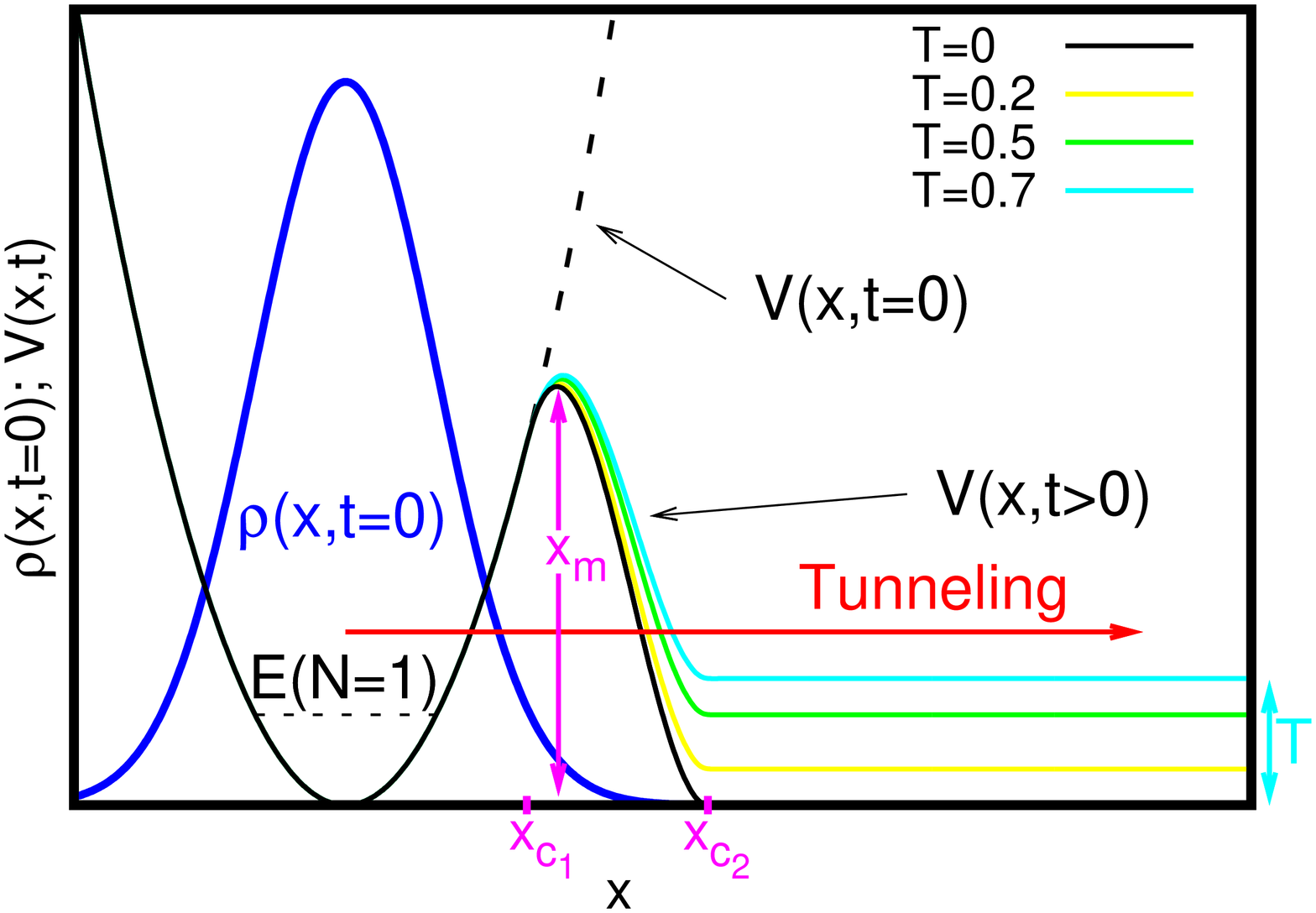}
 \end{center}
\caption{(Color online) Protocol for the tunneling dynamics with non-zero potential threshold.
The initial density (blue line) is prepared as the ground state of the parabolic trap [black dashed $V(x,t=0)$]. Subsequently, the potential is transformed to its open form with a threshold [various solid colored lines, $V(x,t>0)$]. Following this transformation the particles can tunnel to open space. The tunneling process can be controlled by the threshold $T$. The energy of a single, parabolically trapped particle, $E(N=1)$, is indicated by the horizontal black dashed line to guide the eye. In between $x_{c_1}$ and $x_{c_2}$ (indicated by magenta labels on the $x$-axis) the potential is the polynomial $P(x)$ of Eq.~\eqref{potential} with the coefficients as given in Table~\ref{tab1} in Appendix~\ref{polynomial}. All quantities shown are dimensionless.}
\label{pot-tnot0}
\end{figure}

The one-body Hamiltonian $\hat{h}_i$,
\begin{equation}
 \hat{h}_i =  - \frac{1}{2} \partial_{x_i}^2 + V(x_i),
\end{equation}
contains the kinetic energy $- \frac{1}{2} \partial_{x_i}^2$ and the one-body potential $V(x_i)$. Since the scope of the present study is to investigate possible control mechanisms for the many-body tunneling process to open space, the design of the one-body potential $V(x)$ is crucial. It should be a smooth, continuous function which is identical to a parabolic trap in one region of space which is separated from the free, asymptotic part by a barrier. Furthermore, the threshold potential value $T$ in this free, asymptotic region should be easy to modify.
In order to be flexible with the threshold value $T$ it is practical to use a smooth polynomial continuation of the harmonic trap $V_h (x) = \frac{1}{2}x^2$ from $x_{c1} =2$ to $x_{c2}=4$, see Fig.~\ref{pot-tnot0}. The details on the polynomial $P(x)$ are deferred to Appendix~\ref{polynomial}. 
The obtained potential has the form

\begin{widetext}

\begin{equation}\label{potential}
V(x,t)=\left\{
\begin{array}{ll}
\frac{1}{2} x^2 &\qquad t < 0 \\
\Theta(x_{c1}-x)\cdot \frac{1}{2}x^2 + \Theta(x-x_{c1}) \cdot \Theta(x_{c2}-x) \cdot P(x) + \Theta(x-x_{c2})\cdot T &\qquad t \geq 0 
\end{array}
\right.
\end{equation}
 
\end{widetext}

Here, $\Theta(\cdot)$ is the Heaviside step function.
Plots of the potential in Eq.~\eqref{potential} with various values of $T$ are depicted in Fig.~\ref{pot-tnot0}. 

By using a polynomial continuation to the threshold the position of the maximum of the potential, $x_m$, now depends on the threshold $T$ as follows:
\begin{equation}\label{maxbarr}
x_m(T)=2+\frac{1}{3-\frac{3}{4}T}.
\end{equation}
\vspace*{-1.cm}\\
This concludes the exposition of the potential and the Hamiltonian of the system under consideration.
\newpage

\subsection{Quantities of analysis}
To investigate the many-body state and its dynamics it is desireable to have a set of appropriate quantities for the analysis. The full wave function which is available in the MCTDHB computations at any given point in time, is a complicated and high dimensional quantity. It is hence a useful practice to rely on reduced density matrices and their diagonals, i.e., densities, for the purpose of visualization \cite{Glauber1,Glauber,corrglauber,RDMbook,Kaspar}. The reduced one-body density matrix is defined as
\begin{equation}
 \rho^{(1)}(x_1\vert x'_1;t) = N \int \Psi (x_1,...,x_N;t) \Psi^*(x'_1,x_2,...,X_N;t) dx_2 \cdots dx_N.
\end{equation}
When one expands it in its eigenfunctions, the so-called natural orbitals $\lbrace \phi_i(x,t);i=1,M\rbrace$, it takes on the form
\begin{equation}
 \rho^{(1)}(x_1\vert x'_1;t) = \sum_i \rho_i^{(NO)}(t) \phi_i(x_1,t) \phi^*_i(x'_1,t).
\end{equation}
The natural occupations $\rho_i^{(NO)}(t)$, the natural orbitals as well as the diagonal of the reduced one-body density $\rho(x,t)=\rho^{(1)}(x_1=x\vert x'_1=x;t)$, i.e., the so-called density, are very useful quantities to assess quantum many-body dynamics. From the natural occupations one can infer if a system is condensed or fragmented. If only a single macroscopic eigenvalue $\rho_1^{(NO)}$ is present then the system is referred to as condensed \cite{penrose:56}. If multiple eigenvalues $\lbrace \rho_i^{(NO)};i=1,...,M\rbrace$ are macroscopic, then the system is referred to as fragmented \cite{lenzexact2,fragmentothers2,alon.prl2:05,sakmann:09,spekkens:99}.
The density $\rho(x,t)$ describes the probability to find a single particle in the many-body system at a certain position $x$ at a given time $t$.

As a measure of the number of particles inside the parabolic part of the potential (cf. Fig.~\ref{pot-tnot0}), it is instructive to define the nonescape probability, 
\begin{equation}
P^x_{not}(t,T)=\int_{-\infty}^{x_m(T)}\rho(x,t) dx.
\end{equation}
For every propagation with a different threshold one has a different $x_m(T)$, see Eq.~\eqref{maxbarr}.

The coherence of the quantum many-body state can be analyzed with the aid of Glauber's first order normalized correlation function \cite{Glauber1,Glauber,corrglauber,Kaspar}, 
\begin{equation}
g^{(1)}(x'_1,x_1;t)= \frac{\rho^{(1)}(x_1 \vert x'_1;t)  }{ \sqrt{ \rho(x_1,t) \rho(x'_1,t)}}.
\end{equation}
It normalizes the reduced one-body density $\rho^{(1)}$ with its respective diagonal parts $\rho$. If $\vert g^{(1)} \vert^2=1$ holds, the system is coherent. This is only true for the case when $\rho^{(1)}$ is built up as a product of a single complex valued function, cf. Ref.~\cite{Glauber1}. This in turn implies that $\rho^{(1)}$ has only a single eigenvalue $\rho_1^{(NO)}$ and hence $\vert g^{(1)} \vert^2=1$ also means that the system is fully condensed. On the other hand $\vert g^{(1)} \vert^2 < 1$ implies that $\rho^{(1)}$ is built up from several complex valued functions and has several contributing eigenvalues $\lbrace \rho_i^{NO}(t);i=1,...,M\rbrace$. Hence $\vert g^{(1)} \vert^2<1$ also implies a departure from coherence and the occurrence of depletion and eventually fragmentation.

Similarly, the diagonal of the second order normalized correlation function,
\begin{widetext}

\begin{equation}
 g^{(2)}(x'_1=x_1,x'_2=x_2,x_1,x_2;t)=g^{(2)}(x_1,x_2;t)= \frac{\rho^{(2)}(x_1,x_2\vert x'_1=x_1,x'_2=x_2;t)  }{ \sqrt{ \rho(x_1,t) \rho(x'_1=x_1,t) \rho(x_2,t) \rho(x'_2=x_2) }},
\end{equation}
 
\end{widetext}

can be used to infer the second order correlation and coherence of the system under consideration. $g^{(2)}$ normalizes the reduced two-body density $\rho^{(2)}(x_1,x_2 \vert x'_1,x'_2;t)$ with the respective densities $\rho$. $\rho^{(2)}$ is defined as follows \cite{Glauber1,Glauber,corrglauber,Kaspar}:
\begin{equation}
 \rho^{(2)}(x_1,x_2\vert x'_1,x'_2;t) = N (N-1) \int \Psi (x_1,x_2,...,x_N;t) \Psi^*(x'_1,x'_2,x_3,...,x_N;t)dx_3\cdots dx_N.
\end{equation}
Its diagonal is the probability to measure two particles at two positions $x_1,x_2$ simultaneously at a given point in time $t$. From a probabilistic point of view, $g^{(2)}$ can be seen as a measure for the stochastical independence/dependence of the measurement of two particles: If the measurement of the two particles was stochastically independent then $\rho^{(2)}(x_1,x_2\vert x'_1=x_1,x'_2=x_2;t)$ would be equal to $\rho(x_1,t)\rho(x_2,t)$ and $g^{(2)}=\frac{N-1}{N}$ would hold. If, on the contrary, the measurement of the two particles was stochastically \textit{dependent}, then $\rho^{(2)}(x_1,x_2\vert x'_1=x_1,x'_2=x_2;t)$ would be \textit{not} equal to $\rho(x_1,t)\rho(x_2,t)$ and $g^{(2)}=\frac{N-1}{N}$ would \textit{not} hold. Note, that $\frac{N-1}{N} \approx 1$ for $N\gg1$.
The case of $g^{(2)}>1$ is referred to as bunching and the case of $g^{(2)}<1$ as anti-bunching \cite{HBT1,HBT2,HBT3}. From a physical point of view, bunched particles are more likely to reside in two positions together and anti-bunched particles are rather unlikely to reside in two positions together. Quantum fields with no bunching or anti-bunching characteristics, i.e., $g^{(2)}=1$, are referred to as fully second order coherent.

All the above quantities can also be transformed to momentum space, see e.g. Ref.~\cite{Kaspar}. The momentum space representation is versatile to assess quantum dynamics \cite{axel:12} and will be frequently employed throughout the present study. The details on the numerical method, MCTDHB, as well as the computational details can be found in Appendix~\ref{MCTDHB}.

\section{Assembling the many-body process in threshold potentials from basic single-particle processes}\label{modelv2}
\begin{figure}[!]
\begin{center}
 \includegraphics[height=\textwidth,angle=-90]{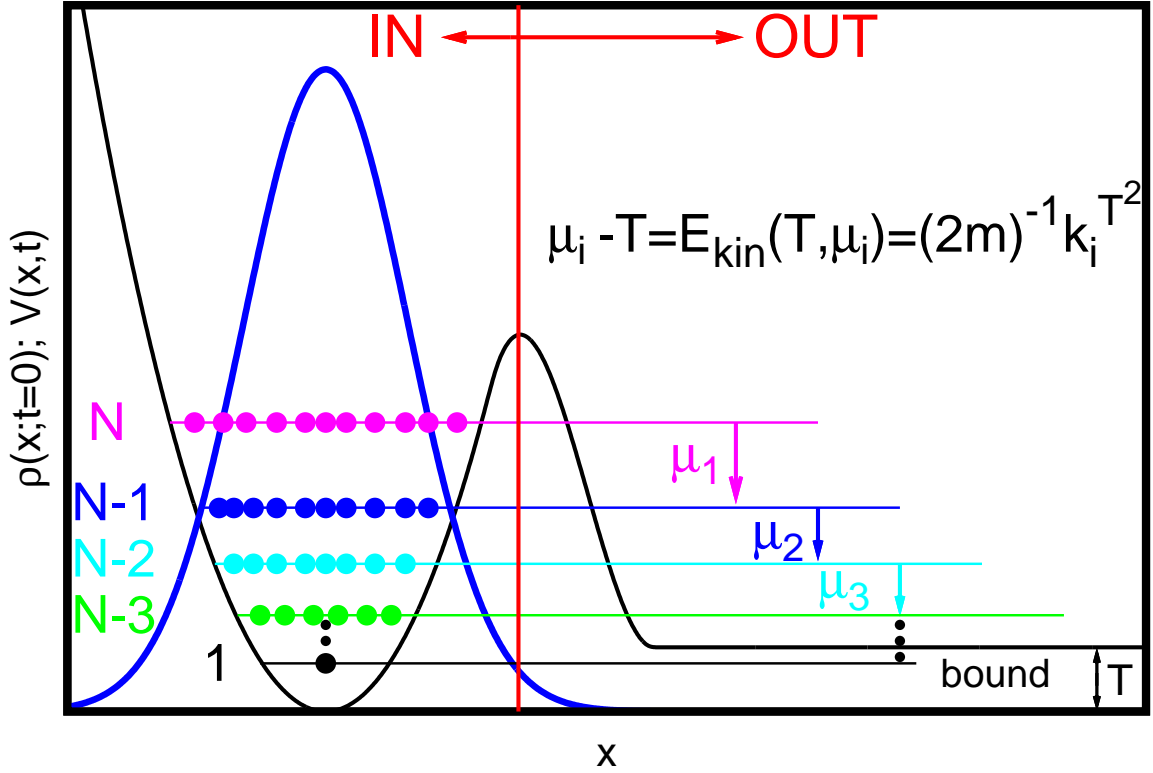}
\end{center}
\caption{(Color online) Static mean-field scheme to model the tunneling processes with a threshold $T$.
The bosons are tunneling from the interior ``IN'' to the exterior ``OUT'' region of space (indicated by the red line). If the threshold $T$ is big enough, some of the states can become bound (see, e.g., the $N=1$ state indicated by the lowest black line). If the state is not bound, the chemical potential $\mu_i$ is first used to overcome the threshold $T$ and thereafter the remainder is transformed to a kinetic energy $E_{kin}(T,\mu_i)$. 
The momenta corresponding to the chemical potentials $k_i=\sqrt{2m(E_{kin}(T,\mu_i))}=\sqrt{2m(\mu_i-T)}; i=N,N-1,...,1$ appear in the momentum distribution, see the arrows in Fig.~\ref{rho_k_t600} and lines in Fig.~\ref{k1_k2_vs_T} below.
All quantities shown are dimensionless.}
 \label{model_mu_T}
\end{figure}

Starting from the model consideration in Ref.~\cite{axel:12} and its successful description of the tunneling dynamics of systems with zero threshold, it is straightforward to adapt the model in order to properly describe the present potentials with a threshold. One can conveniently do that by going through the steps of the model consideration in Ref.~\cite{axel:12} again, taking carefully into account the impact of the threshold onto the energetics -- especially in the exterior part of the potential. As a first step, it is natural to consider the system as split into an ``IN'' part, to the left of the maximum of the barrier at $x_m$, and an ``OUT'' part to the right of the maximum of the barrier at $x_m$. The ``IN'' part is the part of the potential that is classically allowed, i.e., classical particles would be indefinitely confined in the ``IN'' region. For a depiction, see Fig.~\ref{model_mu_T}.
Consider the situation when a single boson has escaped from the ``IN'' to the ``OUT'' region. According to the consideration in Ref.~\cite{axel:12}, the available energy of this boson must come from the energy difference of the trapped systems with $N$ and with $N-1$ particles, $E^N-E^{N-1}=\mu_1$ --  the chemical potential of the $N$-particle system. With this energy available, the ejected boson has to overcome the threshold $T$ -- hence, it remains with an energy $(\mu_1 -T)$ in the ``OUT'' part of the potential to the right of the barrier. As the potential in the ``OUT'' part is flat and the density can be assumed to be small, the ejected boson will convert its available energy to kinetic energy. Analogously, the other particles which are ejected have their available energy from chemical potentials $\mu_i$. As for the system without a threshold, one can hence derive momenta $k_i$ from the related kinetic energies:
 
\begin{widetext}

\begin{equation}
E_{kin}(T,\mu_i)=\mu_i-T=\frac{(k^T_i)^2}{2m} \qquad \Rightarrow \qquad k^T_i=\sqrt{2m(E_{kin}(T,\mu_i)}=\sqrt{2m(\mu_i-T)}. \label{k_threshold}
\end{equation}
 
\end{widetext}

Of course, this assumes that the interaction in the exterior only forces the bosons to occupy different single-particle states and ignores the effect of the interaction on the shape of these states in the ``OUT'' region. It is also evident that in the absence of interaction, all chemical potentials are equal, i.e., $\mu_1=\mu_2=...=\mu_N$.

A particularly interesting feature of the class of potentials with a non-zero asymptotic value is that they can have bound states. If one raises the threshold $T$ beyond the chemical potential $\mu_i$ of a certain parabolically trapped bosonic system then the bosons in the systems with $\mu_i < T$ do not have enough energy to overcome $T$ and thus stay trapped -- hence, the ``IN'' system is in a bound state [cf. Fig.~\ref{model_mu_T} and Eq.~\eqref{k_threshold}]. One can thus control the number of bound particles with both the interaction $\lambda_0$ and the threshold $T$. By manipulating the interaction $\lambda_0$, the energies and especially the chemical potentials can be controlled, and by adjusting the threshold $T$, the number of bound particles can be adjusted. In the case of vanishing interaction, the threshold $T$ controls whether \textit{the whole} system is bound or not. It is convenient to introduce the $\vert N_{IN},N_{OUT} \rangle$ notation, where the first slot of the vector counts the number 
of particles $N_{IN}$ in the ``IN'' 
subsystem and the second slot of the vector counts the number of particles $N_{OUT}$ in the ``OUT'' subsystem according to the partition in Fig~\ref{model_mu_T}. This notation will be used throughout the remainder of the present study. Henceforth, the introduced notation will be referred to as the counting statistics of a given state. The energy for the ``IN'' subsystem, $E_{HO}(N_{IN},\lambda_0)$, is essentially given by the energy of $N_{IN}$ interacting bosons in a parabolic potential. The minimal energy for the ``OUT'' system, $E_{OUT}$, is essentially given by $N_{OUT}$ bosons at rest, i.e., with momentum $k_j=0$, at threshold potential energy, hence, $E_{OUT}=N_{OUT} T$. It follows for the total energy $E_{TOT}$:
\begin{equation}
 E_{TOT}(N_{IN},N_{OUT},T,\lambda_0)=E_{HO}(N_{IN},\lambda_0)+N_{OUT} T. \label{ETOT}
\end{equation}
To summarize, one can adjust the energies of the initial states, $E_{IN}$, by tuning the interaction and the energies of the final states, $E_{OUT}$, by tuning the threshold $T$. The following two sections explore these possibilities for the tunneling bosonic systems with a threshold for $N=2$ and $N=3$ interacting particles, respectively.

\section{Controlling the Dynamics of Two Bosons by the Threshold}\label{cont2}

\subsection{The decay by tunneling dynamics}
\begin{figure}[!]
 \includegraphics[height=\textwidth,angle=-90]{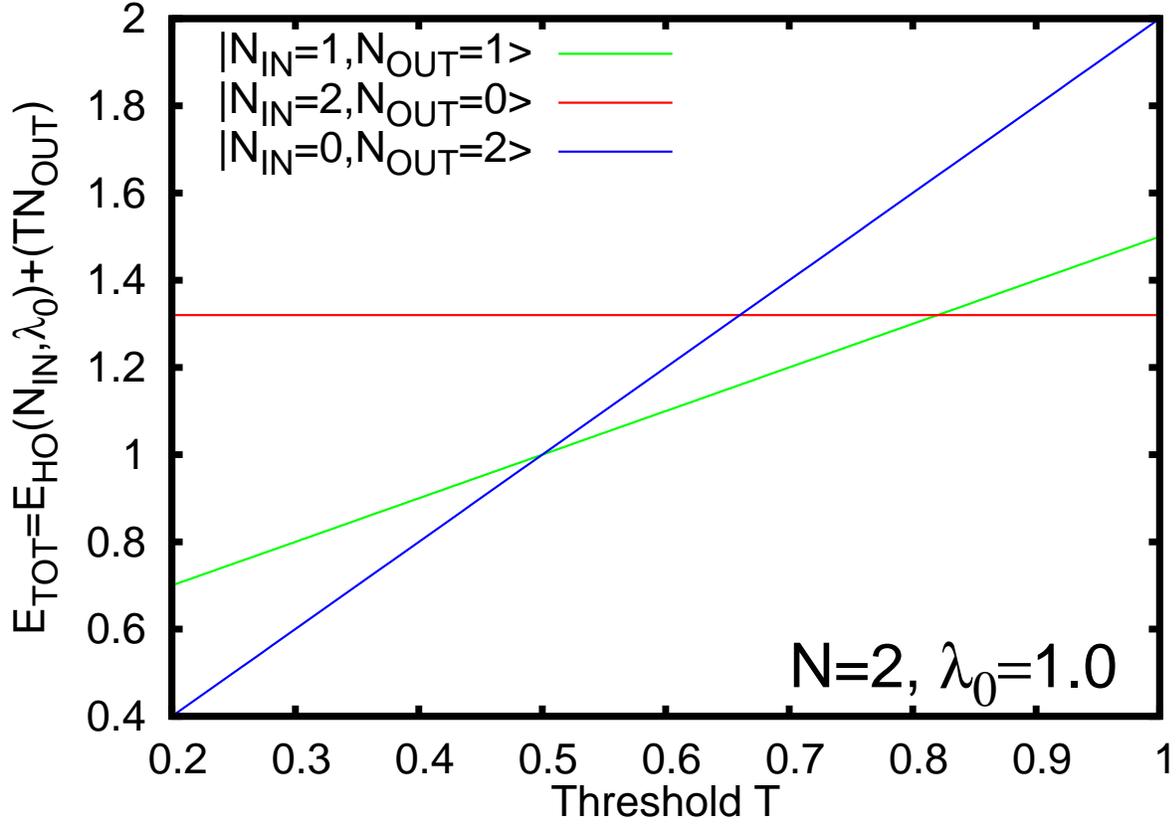}
 \caption{(Color online) Scheme for controlling the two-boson dynamics with the threshold $T$. This plot shows the energies of the possible different final states $\vert N_{IN},N_{OUT} \rangle = \vert 2,0 \rangle, \vert 1,1\rangle$, $\vert 0,2 \rangle$ of two bosons at fixed interaction $\lambda_0=1.0$ with variable threshold $T$. At $T=0.5$, a one-particle bound-state emerges in the trap and at $T\approx0.8$ the two-boson system becomes bound. The crossing points of the energies $E_{TOT}$ determine the (un)availability of final states. See text for further discussion. All quantities shown are dimensionless.} \label{energetics_N2}
\end{figure}

As a first step to explore the dynamics in the new potential with a threshold and the physics of the above model we study the smallest possible many-body system -- two interacting bosons. As we will see, their dynamics is extremely rich. It is instructive to fix first the interaction $\lambda_0=1.0$ and vary the potentials' threshold. Fig.~\ref{energetics_N2} shows the energies of the possible final states with constant interaction and variable threshold for two bosons, i.e., $E_{TOT}(N_{IN},N_{OUT}=2-N_{IN},T,\lambda_0=1.0)$.

\begin{figure}[!]
\includegraphics[height=\textwidth,angle=-90]{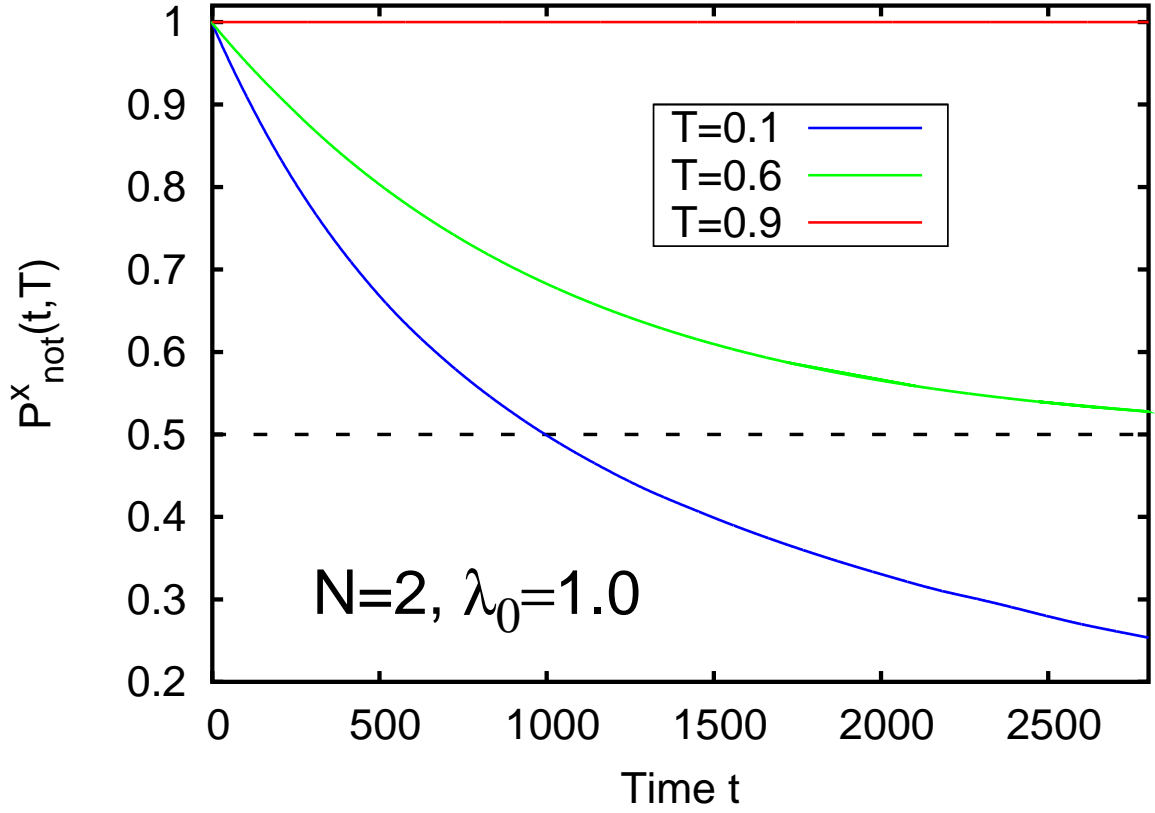}
\caption{(Color online) Nonescape probability for varying thresholds for $N=2$ and $\lambda_0=1.0$.
The nonescape probabilities $P^x_{not}(t,T)$ for different thresholds $T=0.1,0.6,0.9$ is plotted as blue, green and red line, respectively. For $T=0.1$ the final state $\vert 0,2\rangle$, for $T=0.6$ the final state $\vert 1, 1 \rangle$ is favorable. For $T\gtrsim0.8$, the two-boson system is bound, i.e., the only final state available is a bound state for $T=0.9$. The thick horizontal dashed line marks $P^x_{not}=0.5$, the nonescape probability of the final state $\vert 1, 1 \rangle$. See text for further discussion. All quantities shown are dimensionless.}
 \label{T_scan}
\end{figure}

The respective lowest line in Fig.~\ref{energetics_N2} shows the energetically favorable final state for the dynamics. Hence, the crossing points of the lines define critical thresholds at which the energetically favorable final state of the dynamics is changing. One thus would expect that for $T\leq 0.5$ both particles decay, for $0.5<T\lesssim0.8$ one particle decays and one stays bound. For $T\gtrsim 0.8$ the whole system is bound and no particle decays. This behavior is because the final states available are $\vert N_{IN},N_{OUT}=\vert 0,2 \rangle$, $\vert 1,1 \rangle$ and $\vert 2,0 \rangle$, respectively. Since the nonescape probability $P^x_{not}(t,T)$ counts $\frac{N_{IN}}{N}$, it should tend to $0$ for the $\vert N_{IN}=0, N_{OUT}=2 \rangle$ final state, to $0.5$ for the $\vert N_{IN}=1,  N_{OUT}=1 \rangle$ final state and stay at $1$ for the bound $\vert  N_{IN}=2 , N_{OUT}=0 \rangle$ final state. To verify this behavior, Fig.~\ref{T_scan} shows a plot of the nonescape probabilities for the thresholds $T=0.1,0.6$ and $0.9$.

Fig.~\ref{T_scan} shows nicely that the expected behavior of the nonescape probability is recovered and that the prior analysis of the energetics of the problem is applicable. Furthermore, the above analysis demonstrates how the threshold can be used to control the final state of the system by modifying $E_{OUT}=N_{OUT} T$. By tuning $T N_{OUT}$ beyond the (biggest) chemical potential of an $N_{IN}$-body system, one creates an $N_{IN}$-body bound state. This allows for a flexible control of the counting statistics in the ``IN''-subspace \textit{and} the ``OUT''-subspace.

\begin{figure}[!]
\includegraphics[height=\textwidth,angle=-90]{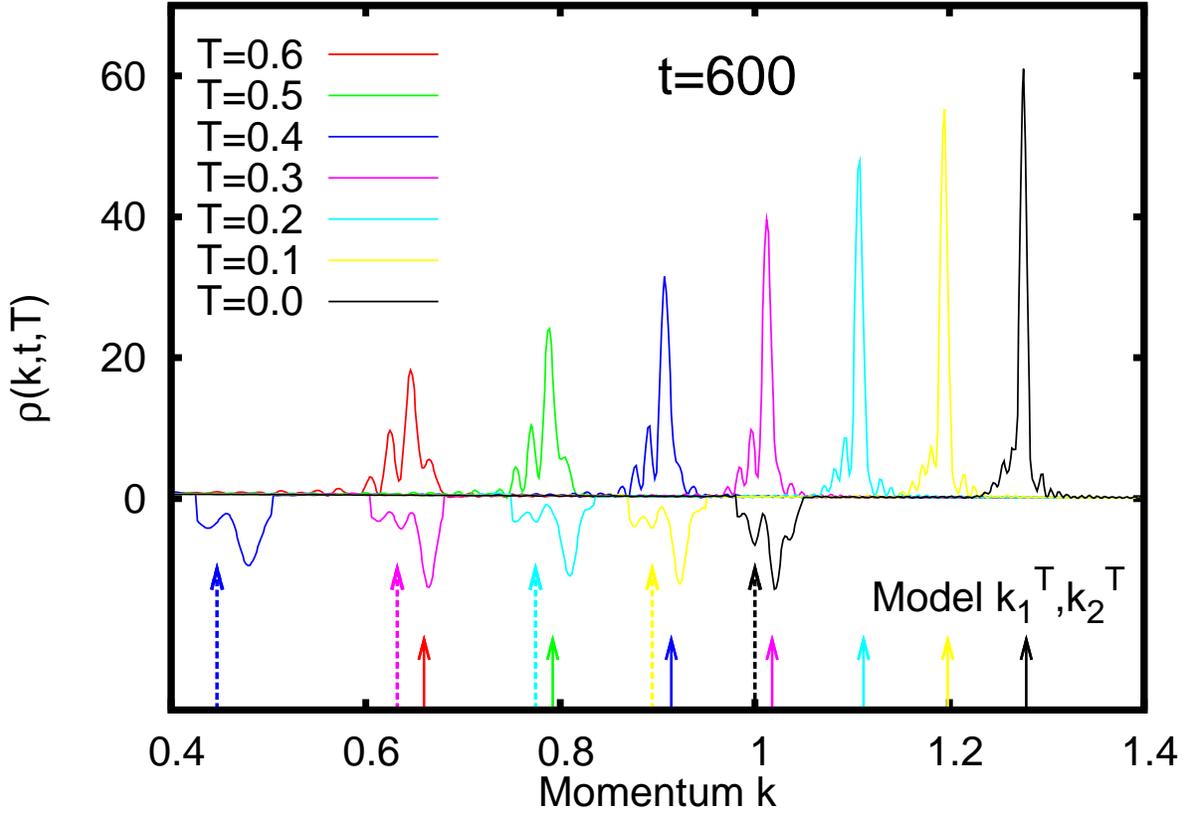}
\caption{(Color online) Effects of various thresholds in the momentum distributions' peak structures of $N=2$ interacting bosons tunneling to open space.
This plot depicts $\rho(k,t=600,T)$ for the tunneling processes in the potentials with thresholds $T=0.0,0.1,...,0.6$. The solid (dashed) line arrows in the bottom of the plot indicate the momenta $k^T_1$ ($k^T_2$) obtained from the model consideration. The momenta are shifted towards $0$ by an increasing threshold $T$. The intensity, i.e., $\rho(k_1^T,t=600,T)$ and $\rho(k_2^T,t=600,T)$, of the peaks is diminished by an increasing threshold. See text for further discussion. All quantities shown are dimensionless.}
\label{rho_k_t600}
\end{figure}

It remains to validate the predictions of the energetics model presented in Fig.~\ref{model_mu_T} on the momenta of the ejected particles, see Eq.~\eqref{k_threshold}. For this validation it is good to inspect a plot of the momentum distributions $\rho(k,t,T)$, see Fig.~\ref{rho_k_t600}. The emitted particles form a peak structure in the momentum distributions. In the case of zero threshold, each emitted particle shows up as a distinct peak in the momentum density in the dynamics, see Ref.~\cite{axel:12}. The momenta of the emitted particles, $k_1,k_2,...$, are essentially time-independent and determined by chemical potentials of systems with decreasing particle numbers $N,N-1,...$ in the case of zero threshold. To assess the effect of the non-zero thresholds on the emission momenta, Fig.~\ref{rho_k_t600} shows a plot of $\rho(k,t,T)$ for $t=600$ and $T=0.1,...,0.6$.

\begin{figure}[!]
\includegraphics[height=\textwidth,angle=-90]{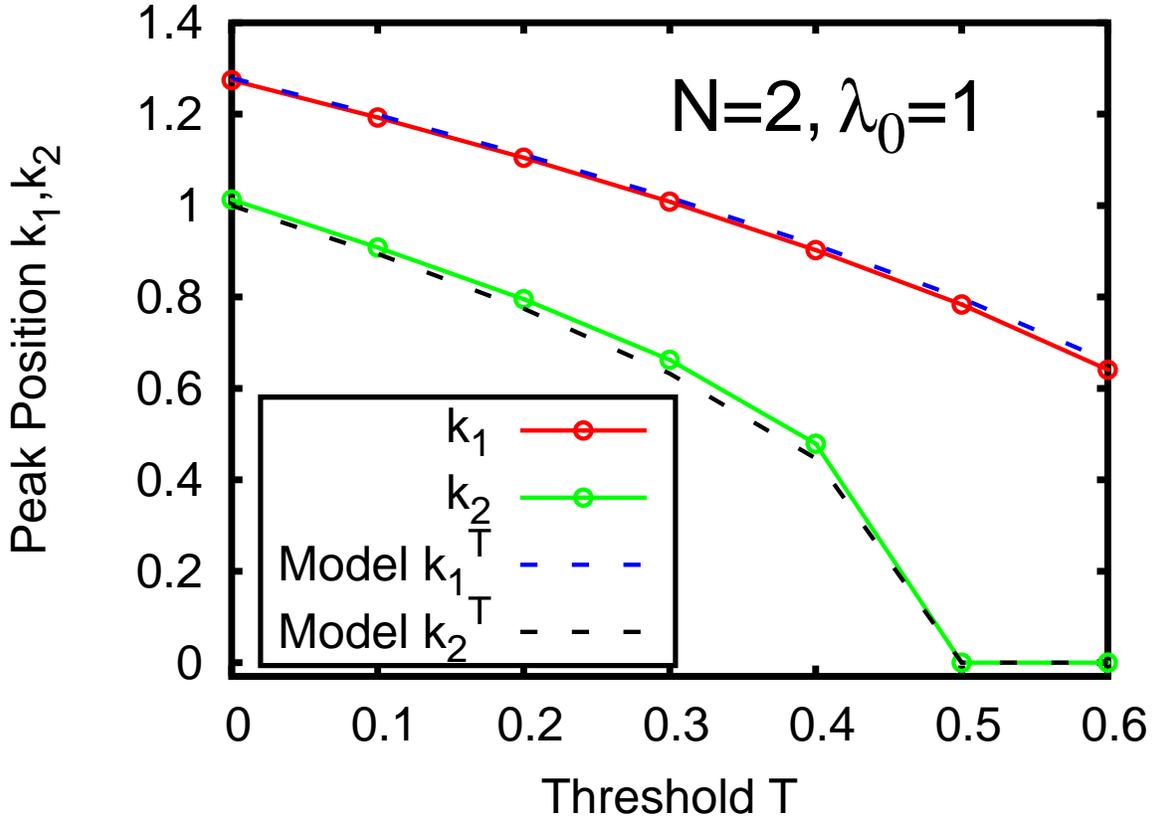}
\caption{(Color online) Comparison of the $N=2$ peak positions to model predictions. The solid red and green lines with points show the peaks' positions $k_1$ and $k_2$ in the exact momentum distributions, the blue and black dashed lines show the model predictions $k^T_1$ and $k^T_2$ from Equation~\eqref{k_threshold}. For the exact solutions circles represent actual data, the lines are drawn to guide the eye. See text for further discussion. All quantities shown are dimensionless.}
 \label{k1_k2_vs_T}
\end{figure}

The changes to the momentum distributions by the threshold are intuitive: The peak structure in the momentum distribution corresponds to the ejected bosons. If the threshold is increased, two effects upon the peaks are seen. First, for a larger threshold each peak is shifted towards $0$, as the escaping bosons have to invest a larger part of their available energy to overcome the higher threshold [cf. Eq.~\eqref{k_threshold}]. Second, the bigger the threshold, the smaller is the intensity of the $k^T_1$ peak, i.e., the peak in the momentum distribution with the largest $k$-value. This means that the increase of the threshold decreases the pace with which the first boson is escaping. A similar reasoning can be applied to the $k^T_2$ peaks and $\rho_(k_2,t=600,T)$.
The agreement of the peaks' positions in $k$-space with the model's prediction is very good (see the arrows in Fig.~\ref{rho_k_t600}). To further assess the validity of the model also for the second peak at $k^T_2$ it is instructive to graph the peak positions' change with varying threshold. This is done in Fig.~\ref{k1_k2_vs_T}.

From the close proximity of the model predictions to the exact solutions' peak positions, see the arrows in Fig.~\ref{rho_k_t600} as well as Fig.~\ref{k1_k2_vs_T}, one can deduce that the tunneling process of the two-boson system can indeed be pictured as an interference of different simultaneous single-boson tunneling processes. These single-boson processes are happening simultaneously. Their momenta are determined by the chemical potentials of systems with different particle numbers. The momenta are shifted by the threshold. When the threshold is above the chemical potential of a certain process a bound state emerges and this process' momentum becomes zero (see $k_2$, i.e., the green line in Fig.~\ref{k1_k2_vs_T} at $T\geq0.5$). The emergence of a bound state in the system closes one of the final states. In the present case of $N=2,\lambda_0=1$ the final state $\vert N_{IN},N_{OUT} \rangle=\vert 0;2 \rangle$ becomes energetically unfavorable for $T\geq0.5$ and consequently the counting statistics of the 
final state are altered to $\vert N_{IN}=1,N_{OUT}=1\rangle$ (see Fig.~\ref{energetics_N2}). Consequently, the nonescape probability $P^x_{not}(t,T)$ of the decay converges to $N_{IN}=1$, i.e. $P^x_{not} \rightarrow 0.5$, from above (see Fig.~\ref{T_scan}).
In summary, the two-body tunneling dynamics to open space can be controlled by the modification of the threshold in the following ways. First, the counting statistics can be controlled with the threshold by creating bound states. Second, the emergence of the bound state can be used as a control on the momentum spectra of the emitted bosons. Peaks can be shifted or even \textit{switched off (on)} by making the corresponding single-boson process energetically unaccessible (accessible).

\subsection{Effect of the Threshold on the Coherence and Correlation Dynamics} 
The dynamics of correlation and coherence in the spirit of Refs.~\cite{Glauber,Glauber1} are of key importance to assess the many-body behavior of the system: Can an effective single particle picture be applied or is the process governed by collective phenomena? 
In the tunneling process of bosonic systems to open space without a threshold, the dynamics of correlations and coherence have shown that collective phenomena occur and have unveiled the mechanism of the tunneling process \cite{axel:12}. The ejected bosons lose the coherence both with the source and among each other. As the processes with a threshold are explained by a similar model, it is interesting to investigate if the correlations or coherence properties are also similar to those in the tunneling process without a threshold. This subsection hence discusses the quantities describing the dynamics of fragmentation coherence: the occupation numbers $\rho_i^{(NO)}(t)$ of the single particle reduced density matrix and the one-particle and two-particle normalized correlation functions $g^{(1)}$ and $g^{(2)}$ \cite{Kaspar}.

\subsubsection{Time-Evolution of the Occupation Numbers}
\begin{figure}[!]
 \includegraphics[angle=-90,width=\textwidth]{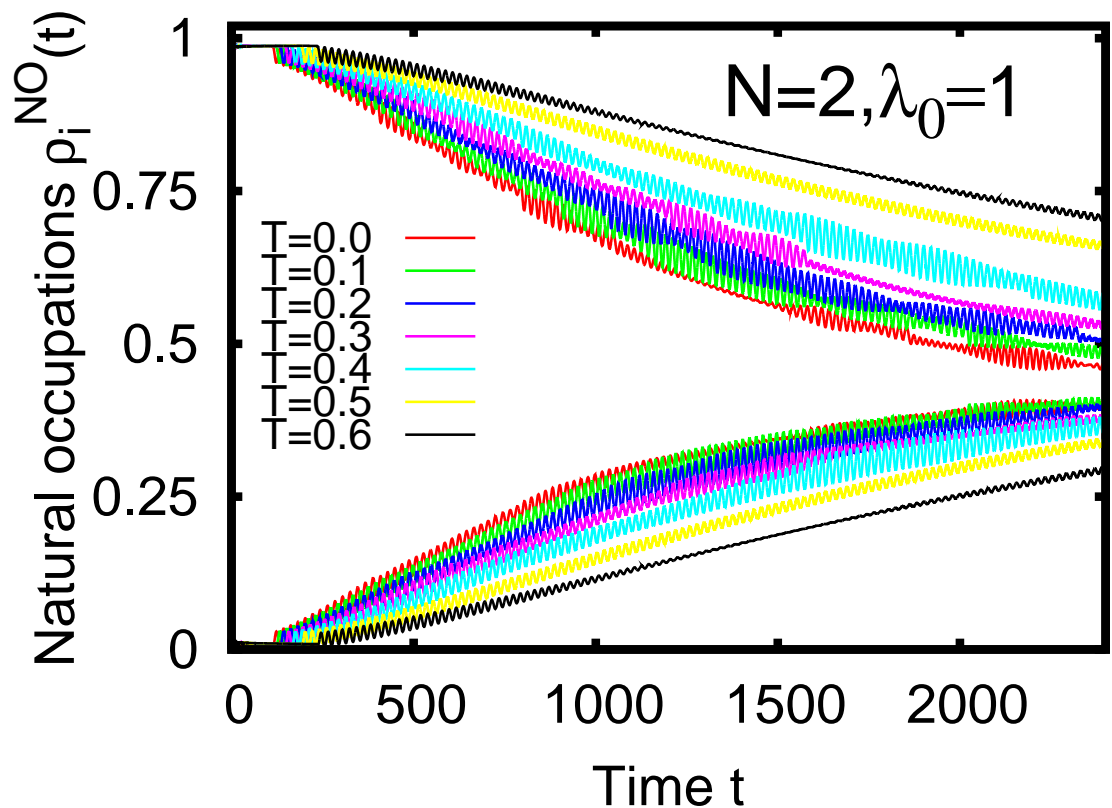}
 \caption{(Color online) Fragmentation is delayed by increasing the value for the threshold. Shown are the time evolutions of the first two occupations, $\rho_1^{(NO)}(t),\rho_2^{(NO)}(t)$ for $N=2$ interacting bosons with $\lambda_0=1.0$ for different thresholds $T$ as various colored solid lines (the threshold $T$ decreases from top to bottom). The occurrence of fragmentation and the buildup of initial depletion are delayed by the increase of the threshold $T$. See text for further discussion. All quantities shown are dimensionless. }
\label{NOCC_T}
\end{figure}

To find the effect of a change in the potential's threshold on the time-evolution of the occupation numbers, it is instructive to plot them for $N=2,\lambda_0=1.0$ and thresholds $T=0.0,0.1,0.2,...,0.6$, as is done in Fig.~\ref{NOCC_T}.

The dependence of the occupation numbers on the increasing thresholds are as follows: as the decay by tunneling process is slowed down by the threshold, also the occurrence of fragmentation is delayed. Furthermore, the initial depletion of the system is delayed, i.e. $\rho_1^{(NO)}\approx 1$ holds for a longer initial time, when $T$ is bigger, see Fig.~\ref{NOCC_T}.

It is very interesting to note that the necessity for a multiconfigurational description  persists also in the cases of $T\geq0.5$ where a one-boson bound state emerges and the counting statistics of the final state are changing from $\vert 0,2 \rangle$ to $\vert 1,1 \rangle$. 
One could naively argue that the final state $\vert 1,1 \rangle$ could be described with a single permanent $\vert n_1,n_2,...,n_M \rangle$. Yet, the chosen $\vert N_{IN},N_{OUT} \rangle$ notation refers to the counting statistics and not to permanents or eigenfunctions of a many-body Hamiltonian. Hence, fragmentation is occurring anyway and one needs many permanents to 
represent the final $\vert N_{IN}=1,N_{OUT}=1 \rangle$ state.  

\subsubsection{Effects of the Threshold on the Correlation Dynamics}
\begin{figure}[!]
\includegraphics[width=\textwidth, angle=-180]{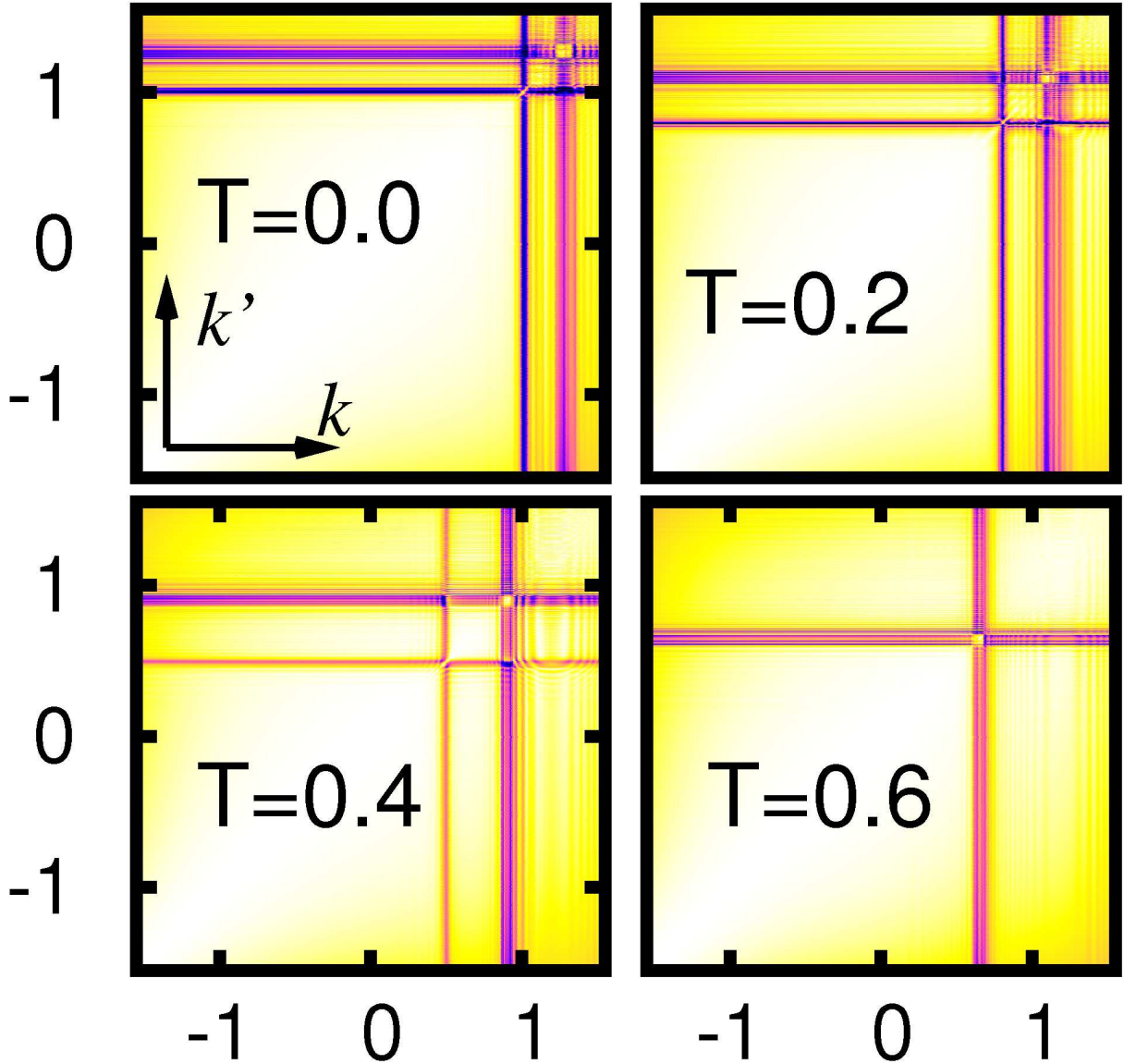}
 \caption{(Color online) Coherence in tunneling to open space of $N=2$ bosons with a threshold.
Shown is the absolute value of the single-particle normalized correlation function, $\vert g^{(1)}(k',k;t)\vert^2$, for $t=600$ for various thresholds $T$. White corresponds to full first-order coherence, i.e., $\vert g^{(1)}\vert^2=1$ and black to full first-order incoherence, i.e., $\vert g^{(1)}\vert^2=0$. The ejected particles lose their coherence with the source. The change of the final state manifests in the absence of a second line where coherence is lost (cf. bottom right plot for $T=0.6$). See text for further discussion. All quantities shown are dimensionless.}
 \label{g1k_vs_T}
\end{figure}

To explore, whether there is an effect of the threshold on the coherence during the fragmentation in the tunneling to open space one has to inspect the normalized single-particle correlation function $g^{(1)}$. In Fig.~\ref{g1k_vs_T} a plot of $g^{(1)}$ in momentum space is given.
In the correlation functions shown in Fig.~\ref{g1k_vs_T}, the single-particle processes from which the many-boson tunneling process is built up are seen as lines of incoherence (the darker lines atop of the white and light yellow background). The positions of these lines coincide with the momenta $k_1,k_2$ predicted by the above model considerations. During the time-evolution the positions of the lines and hence the overall structure of $g^{(1)}$ does not change and it is therefore sufficient to depict $g^{(1)}$ at a single point in time. With the increase of the threshold the system's final state is changed from $\vert 0,2 \rangle$ to $\vert 1,1 \rangle$, i.e., only one of the two particles is decaying to open space for $T\geq0.5$. This change manifests itself in the correlation functions by the \textit{disappearance} of the line at $k_2$ corresponding to the now energetically forbidden process (cf. bottom left and bottom right part of Fig.~\ref{g1k_vs_T}). By the increase of the threshold the loss of coherence 
around the momentum $k_2$ is gradually decreased. Eventually it becomes fully coherent for the $T=0.6$ case, see bottom right panel in Fig.~\ref{g1k_vs_T}. In this manner, peak after peak, corresponding to the model 
processes, the system returns to full coherence, as soon as the respective tunneling channel becomes energetically unfavorable.

\begin{figure}[!]
 \includegraphics[width=\textwidth, angle=-90]{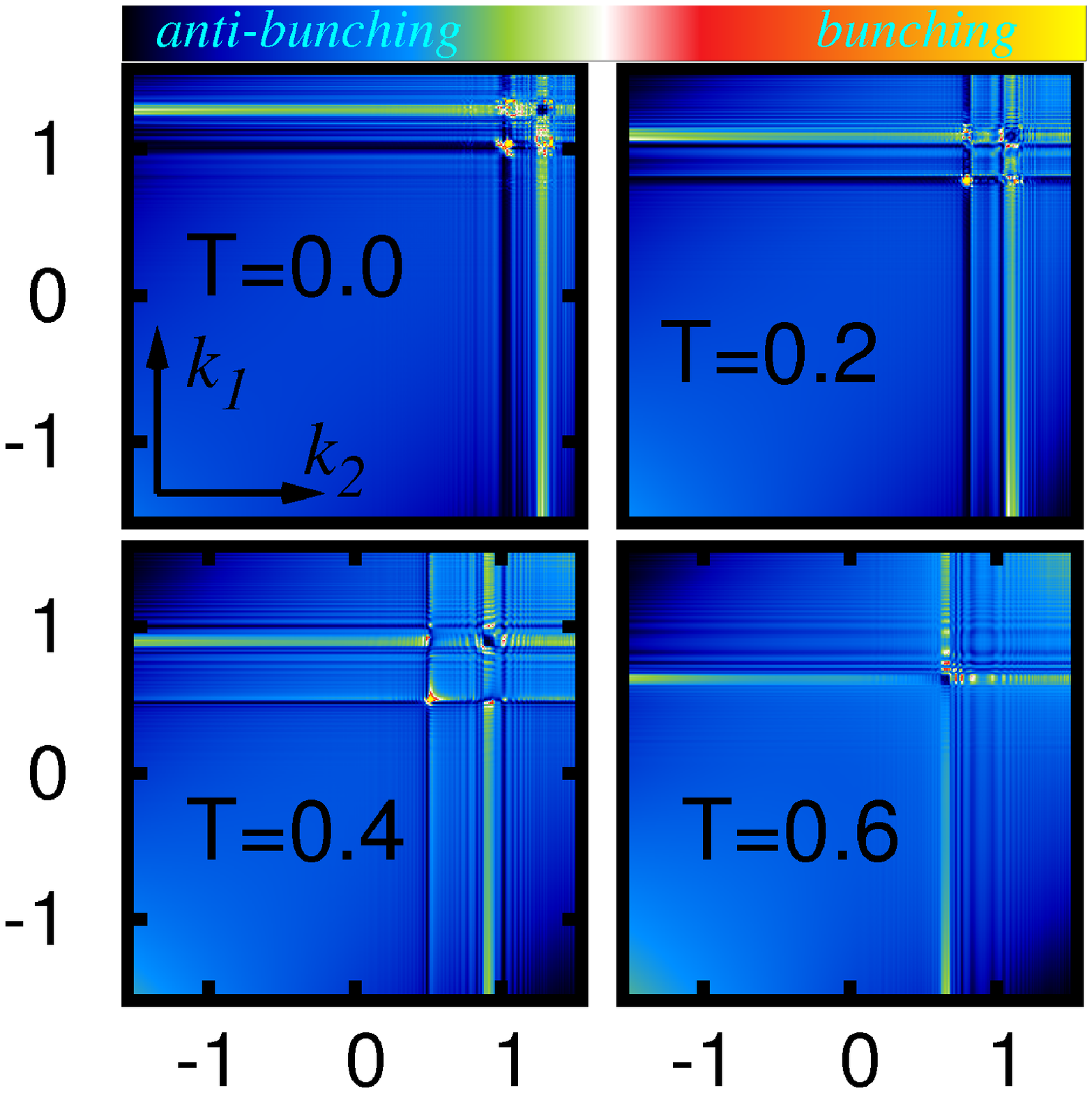}
 \caption{(Color online) Bunching and anti-bunching in tunneling to open space of $N=2$ bosons with a threshold. Shown is the value of the diagonal of the two-particle normalized correlation function, $g^{(2)}(k_1,k_2;t)$ for $t=600$ for various thresholds $T$. The cases of $T=0.0$ and $T=0.2$ show slight anti-bunching for the $k$-space region with the first peak in the momentum distributions and anti-bunching for the region of the second peak. Increasing the threshold gradually switches off the second peak and leaves behind an anti-bunched single line. Throughout the time-evolution (not shown) the peaks on the diagonal at $k_1=k_2=k^T_1$ and $k_1=k_2=k^T_2$ (the diagonal peaks) attain maximal bunching which shifts to the peaks at $k_1=k^T_1;k_2=k^T_2$ and $k_1=k^T_2;k_2=k^T_1$ (the off-diagonal peaks) later on. See text for further discussion. All quantities shown are dimensionless.}
 \label{g2k_vs_T}
\end{figure}

It is interesting that the coherence of the system is also lost in the cases where only a single particle is ejected, see Fig.~\ref{g2k_vs_T}. It is hence instructive to inspect the two-body correlations in the tunneling process. The two-body correlations should show changes of the many-boson process, when the system is switched from two-boson to one-boson decay. In the spirit of Hanbury Brown and Twiss, the situation where $g^{(2)}> 1$ is referred to as bunching and $g^{(2)}< 1$ is referred to as anti-bunching \cite{HBT1,HBT2,HBT3}. For a plot of $g^{(2)}$ in momentum space for $t=600$, see Fig.~\ref{g2k_vs_T}.

The structure of the diagonal of the two-particle normalized momentum correlation function, $g^{(2)}(k_1,k_2;t)$, is intricate: It has a line-structure similar to $g^{(1)}$ in Fig.~\ref{g1k_vs_T}. Yet, in the case of $g^{(2)}$ coherence can be lost in two ways -- through bunching, i.e. $g^{(2)}>1$, or anti-bunching, i.e. $g^{(2)}<1$. In the case of bunching of the two momenta $k_a$ and $k_b$, the two particles are more likely to have these two momenta and in the case of anti-bunching it is rather unlikely that the two particles have these momenta at the same time. From the general structure of Fig.~\ref{g2k_vs_T} one can read that the part of the cloud which is at rest, i.e., where $k_1=k_2\approx 0$, is initially and throughout the tunneling process a slightly anti-bunched, almost second order coherent entity, because $g^{(2)}\lesssim 1$ holds for $k_1=k_2\approx 0$ at all different thresholds $T$. The lines are located at $k_1^T$ and $k_2^T$ where the peaks in the momentum distribution are. The first line 
at the larger $k_1^T$ 
shows bunching whereas the second line at the smaller $k_2^T$ shows anti-bunching. This means that it is likely 
to find one boson at rest and one with $k_1^T$, while it is rather likely that the second boson also propagates when one finds the first one at $k_2^T$. The change in the final state of the tunneling process which is due to the energetics as explained in Fig.~\ref{energetics_N2} is again visible by the disappearance of the line around $k_2^T$ (cf. bottom right part of Fig.~\ref{g2k_vs_T}). The diagonal point at $k_1^T=k_1=k_2$ is strictly anti-bunched in this case. The line structure, see bottom right part of Fig.~\ref{g2k_vs_T}, is strictly bunched. This behavior is expected, because only one boson can leave and hence it is becoming more and more likely to have a boson at rest and another one propagating with $k_1^T$.

In the cases where both bosons are decaying the degree of sequentiality can be assessed with $g^{(2)}$: When one analyzes the line around $k_2^T$ in the plots of $g^{(2)}$ for $T=0.0,0.2$ and $0.4$, bunching occurs only at the intersections with the other lines -- this means that it is very likely that one boson has already left the trap and 
propagates with momentum $k^T_1$ when the second one follows with 
momentum $k^T_2$. In the case of the two-particle decay, the intensities of the peaks at the intersection points of the lines are time-dependent, see Fig.~\ref{g2k_vs_T}. Initially, the peak at $k_1=k_2=k^T_1$ is dominant, followed by bunching also in the $k_1=k_2=k^T_2$ and off-diagonal $k_1=k_1^T;k_2=k_2^T$, and $k_1=k_2^T;k_2=k_1^T$ regions. Finally the off-diagonal peaks will become dominant, owing to the fact that the final state of the dynamics contains two bosons in the ``OUT'' region, each propagating with a specific momentum.

To summarize, the dynamics of the tunneling process to open space can be managed by the threshold $T$. The occurrence of bound states manifests itself by closing final states of the dynamics. The momenta in the decay process are obtained from the chemical potentials of systems with reduced particle number and the threshold. The coherence of the ejected particles with the source is lost and the bunching and anti-bunching properties explain to which degree the processes occur (non-)sequentially. The dynamics of the correlation functions $g^{(1)}$ and $g^{(2)}$ can be managed side by side with the momentum distributions with the threshold.

\section{Controlling the Dynamics of Three Bosons by the Interactions}\label{cont3}

\subsection{Energies of the final states of the dynamics}
\begin{figure}[!]
\includegraphics[height=\textwidth,angle=-90]{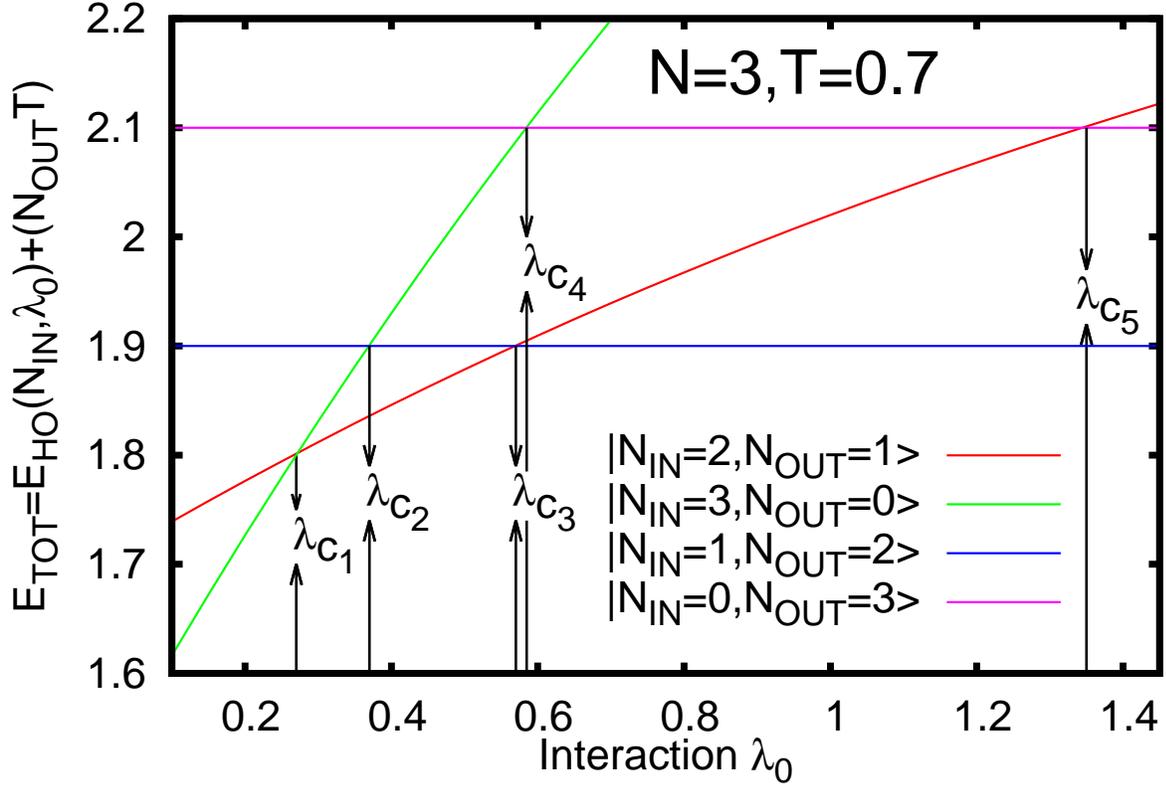}
\caption{(Color online) Energetics of the $N=3$ system and a threshold of $T=0.7$ with variable interaction strength $\lambda_0$.
This plot shows the minimal energies $E_{TOT}(N_{IN},N_{OUT},T,\lambda_0)$ needed to allow the different final states $\vert N_{IN},N_{OUT} \rangle = \vert 3,0 \rangle, \vert 2,1\rangle$, $\vert 1,2 \rangle$, and $\vert 0,3 \rangle$, as solid green, red, blue, and magenta line, respectively. The crossing points at $\lambda_{c_k}$ show for which interactions the particular final states are available. See text for discussion. All quantities shown are dimensionless.}
\label{energetics_N3}
\end{figure}

The aim of the present and subsequent sections is to underline and corroborate the generality of the findings of the previous section on $N=2$ bosons also for bigger particle numbers. The control mechanism employed for the final states is generalized in this section: Instead of the previous control by the threshold $T$ of the $N=2$ tunneling dynamics, the interaction strength $\lambda_0$ is now used and the threshold is kept at a fixed value $T=0.7$. By changing $\lambda_0$ one can determine which final states are favorable in the dynamics.
To start, a discussion of the energies of the available $\vert N_{IN},N_{OUT}\rangle$ states in the $N=3$ tunneling dynamics is appropriate. 
A plot of $E_{TOT}(N_{IN},N_{OUT},T=0.7,\lambda_0)$ for $N=N_{IN}+N_{OUT}=3$ particles is given in Fig.~\ref{energetics_N3}.

The four possible final states' energies have a generally different dependence on the interaction and one finds $5$ intersections in Fig.~\ref{energetics_N3} that correspond to critical interaction strengths that are labeled by $\lambda_{c_i},i=1,...,5$. 
The energies of $\vert 3,0\rangle$ and $\vert 2,1 \rangle$, i.e., $E_{TOT}(N_{IN}=3,N_{OUT}=0,T=0.7,\lambda_0)$, and $E_{TOT}(N_{IN}=2,N_{OUT}=1,T=0.7,\lambda_0)$, are dependent on the strength of the interaction, $\lambda_0$, while the energies of both $\vert 1,2 \rangle$, and $\vert 0,3 \rangle$ are independent of $\lambda_0$ in the considered model, see Eq.\ref{ETOT} and Fig.~\ref{energetics_N3}. The reason is that the energy of a single boson does not depend on the interaction [$E_{HO}(1,\lambda_0)=0.5$] and the interaction of the emitted bosons in the exterior is neglected in the model. The energy of $\vert 1,2 \rangle$ is $E_{TOT}=(1,2,0.7,\lambda_0)=0.5+2\cdot T=0.5+1.4=1.9$ and the energy of $\vert 0,3 \rangle$ is $E_{TOT}(0,3,0.7,\lambda_0)=E_{HO}(N_{IN}=0,\lambda_0)+3\cdot T=3\cdot0.7=2.1$. These are the minimal energies the system needs in order to eject two or three particles, respectively. The energy of the final state, in which a single particle has tunneled, $E_{TOT}(2,1,0.7,\lambda_0)=E_{HO}(
N_{IN}=2,\lambda_0)+1\cdot T=0.7+E_{HO}(N_{IN}=2,\lambda_0)$, is dependent on the interaction, because the energy of the trapped system, $E_{HO}(N_{IN}=2,\lambda_0)$, with two bosons depends on the interaction and so does the energy of the trapped system $\vert 3,0\rangle$ for the analogous reason.

If one hence chooses an interaction $\lambda_0$ smaller than $\lambda_{c_1}$ for the initial state of $N=3$ parabolically trapped particles, the system is bound with a threshold of $T=0.7$ because the possible final states, $\vert N_{IN}=2, N_{OUT}=1 \rangle$, $\vert N_{IN}=1, N_{OUT}=2 \rangle$, and $\vert N_{IN}= 0, N_{OUT}=3 \rangle$ are energetically not available. If one chooses an interaction of $\lambda_{c_2} > \lambda_0 > \lambda_{c_1}$ then the final state $\vert N_{IN}=2, N_{OUT}=1\rangle$ is energetically allowed, i.e., $E_{HO}(3,\lambda_0)>E_{TOT}(2,1,0.7,\lambda_0)$, but the other final states are energetically forbidden. In this regime the $N=3$ system should thus decay by emitting a single boson, leaving behind two bound bosons. In the case of e.g. $\lambda_{c_3}>\lambda_0>\lambda_{c_2}$ two final states, i.e., $\vert 2,1\rangle$, and $\vert 1,2\rangle$, are energetically allowed, because $E_{HO}(3,\lambda_0)>E_{TOT}(1,2,0.7,\lambda_0)>E_{TOT}(2,1,0.7,\lambda_0)$. In this situation it turns out that the energetically lowest configuration is the actual final 
state. This 
means that, e.g., in the above case of $\lambda_{c_3}>\lambda_0>\lambda_{c_2}$ one finds the final state of the dynamics to be $\vert 2,1 \rangle$, i.e., the ejection of a single particle is preferred. This can be explained intuitively by the physics of decay processes: The rate at which a decay process is occurring is determined by the overlap of the initial and the final states. For the following discussion, it is appropriate to remind the reader that the $\vert N_{IN}, N_{OUT} \rangle$ notation does not correspond to Fock states and in this notation states that have different occupations can hence have an overlap.  Intuitively, the overlap of the $\vert N_{IN}=3,N_{OUT}=0\rangle$ and  $\vert N_{IN}=2,N_{OUT}=1\rangle$ states is bigger than that of the $\vert N_{IN}=3,N_{OUT}=0\rangle$ and $\vert N_{IN}=1,N_{OUT}=2\rangle$ states. This is simply due to their contributions in the ``IN'' subspace. Furthermore, there is also an overlap of the $\vert N_{IN}=1,N_{OUT}=2\rangle$ and the $\vert N_{IN}=2, N_{OUT}=1\rangle$ states. This means that there is a rate with which $\vert N_{IN}=1,N_{OUT}=2\rangle$ is transformed to $\vert N_{IN}=2,N_{OUT}=1\rangle$. With this reasoning the final state is hence the energetically lowest final configuration. One can apply a similar reasoning for the other critical interactions $\lambda_{c_4},\lambda_{c_5}$. It is interesting to note the peculiarity of the process -- determined by the overlap of  $\vert N_{IN}=1,N_{OUT}=2\rangle$ and $\vert N_{IN}=2,N_{OUT}=1\rangle$: the trapped particle number $N_{IN}$ is actually increasing by one. With this reasoning it should thus be possible to find sets of parameters for which the system's nonescape probability is increasing for a limited amount of time. This is at times at which the predominant part of the wave function is similar to, e.g., $\vert N_{IN}=1,N_{OUT}=2\rangle$. In all the presented examples in this section this was not the case. This makes the conclusion tempting, that the observation of this counterintuitive regime is not 
possible because the above-mentioned transformation of $\vert N_{IN}=1,N_{OUT}=2\rangle$ to $\vert N_{IN}=2,N_{OUT}=1\rangle$ is very efficient. Therefore, any population in $\vert N_{IN}=1,N_{OUT}=2 \rangle$ is momentarily shifted 
to $\vert N_{IN}=2,N_{OUT}=1\rangle$ and the corresponding counting statistics cannot be found. % In order to consistently investigate such populations so-called loss operators -- projection operators onto a certain distribution of particles in a partitioned Hilbert space -- are needed. With these loss operators the time-evolution of the population of any final state is in principle accessible. In the current implementation of the MCTDHB\cite{streltsov2010}, these loss operators are not yet available. 
To assess the validity of the above considerations, it remains to quantify the counting statistics in the dynamics with the density-related nonescape probabilities $P^x_{not}(t,T)$.

\subsection{Decay by tunneling dynamics}

\begin{figure}[!]
\includegraphics[height=\textwidth,angle=-90]{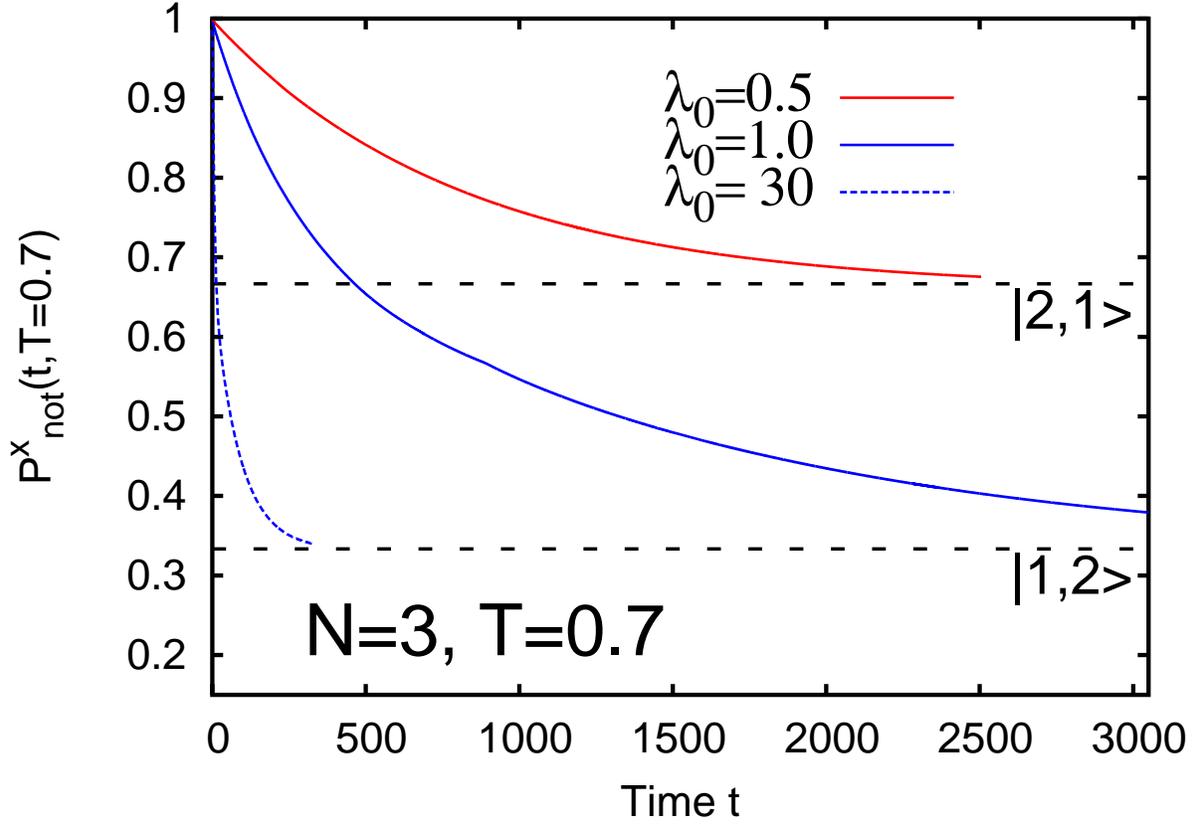}
\caption{(Color online) Time-evolution of the nonescape probabilities for different final states of $N=3$ bosons tunneling to open space. 
The dependence on the interaction $\lambda_0$ of the nonescape probability in a potential with a threshold of $T=0.7$ is shown. The interactions and the threshold were chosen according to the energetics (cf. Fig.~\ref{energetics_N3}) of the process such that there is a bound state for either $N_{IN}=1$ (blue solid and dashed lines) or $N_{IN}=2$ (red solid line) bosons. The two dashed horizontal lines are at $P^x_{not}=\frac23$ and $P^x_{not}=\frac13$, respectively. See text for further discussion. All quantities shown are dimensionless.}
\label{Pnot_N3}
\end{figure}

To verify the above the time-evolution of the nonescape probabilities, $P^x_{not}(t,T)$ in the given example of $N=3$ bosons in a potential with $T=0.7$ for various interactions $\lambda_0$ are analyzed. For a plot of the nonescape probabilities, corresponding to the different possible final states, see Fig.~\ref{Pnot_N3}.

Also for $N=3$ the behavior of the nonescape probabilities is as predicted from the energies and the availability of the final states: when a certain final state becomes energetically unavailable, then the counting statistics of the final state change. For example, for $\lambda_0=0.5$, the lowest available final state in the $\vert N_{IN},N_{OUT}\rangle$ notation is $\vert 2, 1 \rangle$ -- consequently, the norm of the density in the ``IN'' subspace, i.e., the nonescape probability $P^x_{not}(t,T)$, converges to $\frac{2}{3}$. In the case of the stronger interaction $\lambda_0=1.0$, the final state $\vert 1, 2 \rangle$ is energetically favorable and consequently the nonescape probability converges to $\frac{1}{3}$. The model introduced in Section~\ref{modelv2} is indeed accurately predicting the counting statistics of the tunneling to open space process with a threshold. To corroborate this finding and to assess the generality of the model also for stronger interactions, Fig.~\ref{Pnot_N3} shows also a plot of the nonescape probability for the very strong interaction $\lambda_0=30.0$. For such a 
strong interaction, the initial state is fermionized and one would expect that the model description would be inaccurate if its validity depended on the inter-particle interactions. Yet, the model prediction of a nonescape probability $P^x_{not}(t,T)$ of $\frac{1}{3}$ for the final state  $\vert 1, 2 \rangle$ still holds. Of course, the decay happens at a much faster pace in this stronger interacting case. Hence, one might expect that the model consideration should hold for particle numbers $N>3$ and for general interaction strengths. In order to asses this generality of the model, section \ref{contmany} discusses the control of the tunneling dynamics of $N=101$ bosons. 

\subsection{Coherence and Correlations in the Tunneling Process with a Threshold of N=3 bosons}
\begin{widetext}

\begin{figure}[!]
\includegraphics[height=\textwidth, angle=-90]{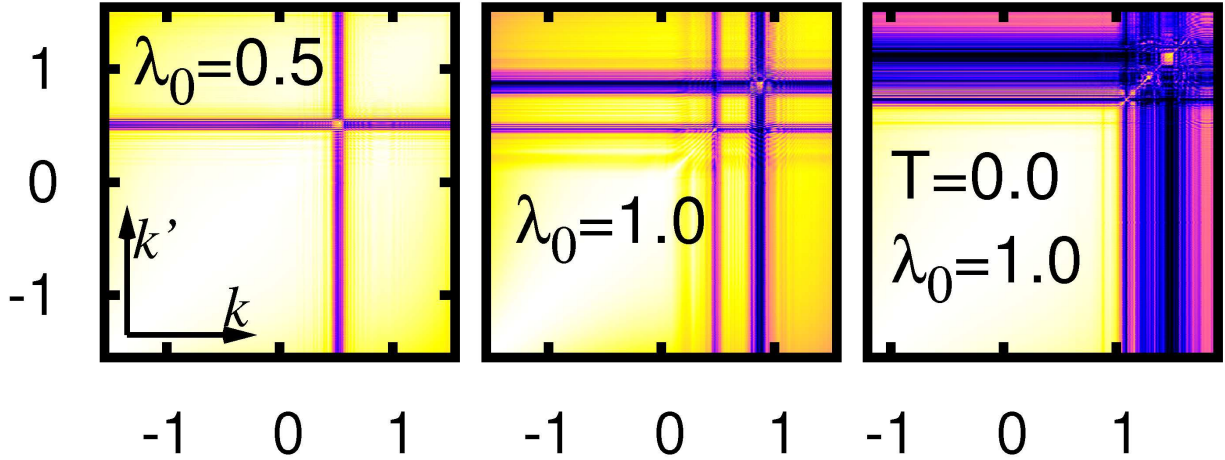}
\caption{(Color online) Coherence dynamics in tunneling to open space with a threshold for $N=3$ bosons. The first order correlation function $\vert g^{(1)} \vert^2$ is plotted for three different final states with $N_{IN}=2$, $1$, and $N_{IN}=0$ in the left, middle and right panel, respectively for the time $t=800$. White corresponds to $\vert g^{(1)}\vert^2 =1$ and black to $\vert g^{(1)}\vert^2=0$. The structure of $\vert g^{(1)}\vert^2$ is determined by the peak structure in the momentum distribution. The left panel, for the interaction $\lambda_0=0.5$ and threshold $T=0.7$, shows a single line at the momentum with which the single boson escapes. In the middle panel, for $\lambda_0=1.0$ and $T=1.0$, two bosons are emitted and the wave function looses its coherence at precisely their respective momenta. For convenience, the right panel shows the case of $\lambda_0=1.0$ and $T=0.7$ where all $N=3$ bosons can decay -- and consequently $3$ lines show up where $\vert g^{(1)}\vert^2\approx 0$ at the momenta $k_
1,k_2,k_3$ of the three bosons that escape (note the three minima in the right panel on the right and top borders for $k>1.0$,$k'>1.0$). These momenta are predicted accurately by the model [cf. Eq.~\eqref{k_threshold}]. See text for further discussion. 
All quantities shown are dimensionless.}
\label{corr1_T_N3}
\end{figure}

\end{widetext}

In order to assess the effects in the processes' correlation and coherence dynamics it is best to take a look at the correlation functions $g^{(1)}$ and $g^{(2)}$ in momentum space. 
Fig.~\ref{corr1_T_N3} shows the $\vert g^{(1)}(k_1,k'_1,t=800)\vert^2$  for varying interactions and fixed threshold $T=0.7$ in the left and center panels. For convenience and in order to display all possible final states of the dynamics with an $\vert N_{IN}=3,N_{OUT=0} \rangle$ initial state, the right panel of Fig.~\ref{corr1_T_N3} shows the coherence in the $T=0$ dynamics.
 
Indeed, the behavior of the case of $N=2$ bosons is reproduced in the dynamics of the coherence in the tunneling to open space process of $N=3$ bosons. By increasing the interaction $\lambda_0$ across the critical value for the availability of a certain final state, new lines which are incoherent with the source at rest and among each other show up (cf. left and middle panel of Fig.~\ref{corr1_T_N3}). Side by side with the momentum distributions, the first order coherence in the process can hence also be managed by the manipulation of $\lambda_0$. Of course, the dynamics shown involve the fragmentation of the initially coherent sample of $N=3$ parabolically trapped bosons. The time evolution of the occupation numbers and the momentum distributions in this case resembles the one in Figs~\ref{NOCC_T} and \ref{rho_k_t600} and is not shown, therefore.

\begin{widetext}

\begin{figure}[!]
\includegraphics[height=\textwidth, angle=-90]{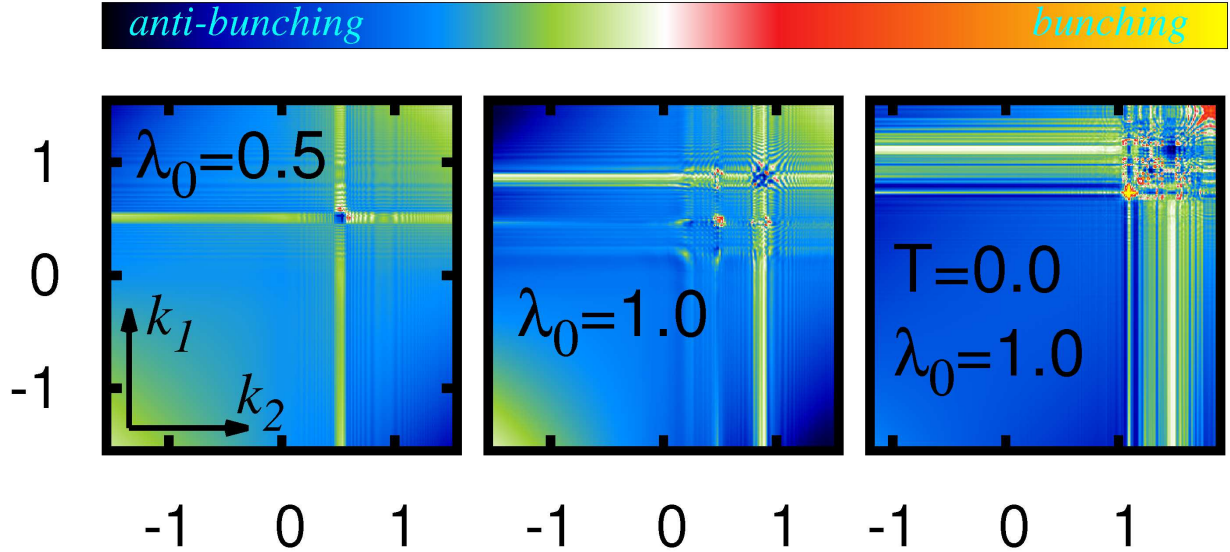}
\caption{(Color online) Bunching dynamics of tunneling to open space with a threshold for $N=3$ bosons. The second order correlation function $g^{(2)}(k_1,k_2;t=600)$ is plotted for three different final states with $N_{IN}=2$, $1$, and $N_{IN}=0$ in the left, middle and right panel, respectively for the time $t=600$. The line structure of the first order correlation functions in Fig.~\ref{corr1_T_N3} is almost preserved. The lines corresponding to the biggest momenta show slight anti-bunching, i.e., $g^{(2)}<1$ . In their crossings the anti-bunching intensifies $g^{(2)}\approx 0$. Where the lines corresponding to the bigger $k$ intersect the lines of the smaller momenta bunching, i.e., $g^{(2)}>1$ occurs -- this corresponds to the sequential ejection of two bosons, see top right part of middle and right panel. See text for further discussion. All quantities shown are dimensionless.}
\label{corr2_T_N3}
\end{figure}

\end{widetext}

It remains to find out what are the two-body properties of the process. For this purpose, a plot of the second order coherence $g^{(2)}$ is shown in Fig.~\ref{corr2_T_N3}.
 
The structure of the first order coherence $\vert g^{(1)}\vert^2$ in Fig.~\ref{corr1_T_N3} is preserved for the diagonal part of $g^{(2)}$ in Fig.~\ref{corr2_T_N3}. The anticipated behavior from the case of $N=2$ bosons (cf. Fig.~\ref{g2k_vs_T}) prevails: the bunching is mainly for the off-diagonal intersections of the slightly anti-bunching lines at the different momenta. The degree of the bunching on the diagonal and on the off-diagonal shows the sequentiality of the process. For example, the line corresponding to the biggest momentum is always the closest to coherent (i.e., white in Fig.~\ref{corr2_T_N3}) and the anti-bunching for this line on the diagonal is the strongest. Hence, the boson which is emitted and is propagating at the corresponding momentum $k_1^T$ is very unlikely to be found, if another boson also propagates and the same momentum. Furthermore, the boson propagating with $k_1^T$ is coherent, i.e., uncorrelated with all the other momenta -- one could say that it does not feel the remainder 
of the 
$3$-boson system. This explains the good applicability of the model introduced in Section~\ref{modelv2}. While the escaped bosons lose their first order coherence with the source, their second order coherence is preserved. This is because the process accounting for a single line in the $g^{(1)}$ or $g^{(2)}$, respectively, is  a single-particle process. The model's elementary processes describe exactly such a behavior. Similar reasoning can be applied to the other lines in $g^{(2)}$. This concludes the discussion of the first- and second-order coherence in the three-boson process of tunneling to open space.

\section{Controlling the Many-Body Process}\label{contmany}
\begin{widetext}

\begin{figure}[!]
\includegraphics[angle=-90,width=\textwidth]{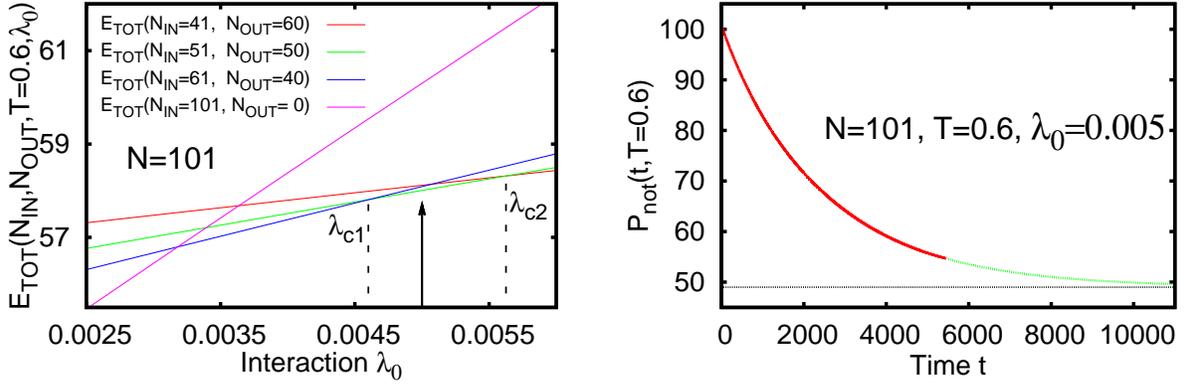}
\caption{(Color online) Energetics and nonescape probability for the tunneling to open space of $N=101$ bosons. 
Left Panel: Shown are the energies of the final states $\vert N_{IN},N_{OUT}\rangle=\vert 41,60\rangle, \vert 51,50 \rangle$, and $\vert 61, 40 \rangle$. When one tunes the interaction $\lambda_0$ such that it is in between the crossing points (marked by the black dashed vertical lines) of the green and the red and magenta solid lines at $\lambda_0=\lambda_{c1}$ and $\lambda_0=\lambda_{c2}$, the energetically most favorable state will be with $N_{IN} \in (41,61)$ and $N_{OUT}=N-N_{IN}$ particles. The black arrow shows the interaction $\lambda_0=0.005$ chosen for the propagation. Right Panel: Shown is the nonescape probability $P^x_{not}(t,T)$ of $N=101$ particles with $\lambda_0=0.005$ (red solid line). According to a least squares fit (see green, dashed line), the final state is $\vert N_{IN}=49,N_{OUT}=52\rangle$ and hence in the range assessed from the energetics in the left panel. See text for further discussion. All quantities shown are dimensionless.}
\label{Pnot_N101}
\end{figure}
 
\end{widetext}

In this section, the control schemes using the threshold $T$ and the interparticle interaction $\lambda_0$ found in the previous sections are applied to the dynamics of $N=101$ bosons. The intention is to obtain a desired final state, of roughly $N_{IN}=50$ bosons. The strategy chosen is to first fix a threshold and then tune the interaction appropriately -- it is noteworthy that first choosing an interaction and thereafter adjusting the threshold is also possible.
Fig.~\ref{Pnot_N101} shows the energetics and nonescape probability for $N=101$ particles. In this case the threshold was fixed to $T=0.6$. One can tune the tunneling process' counting statistics by modifying the interactions in order to obtain an $N_{IN}\approx 50$ bound state by the presented reasoning.

Fig.~\ref{Pnot_N101} validates the model [cf. Section \ref{modelv2} and Eqs.~\eqref{k_threshold},\eqref{ETOT}] for a general number of particles. This makes the formulation of a protocol for the deterministic production of a desired $N$-boson state possible. In the case of a fixed potential threshold one can tune the interactions $\lambda_0$ such that the energy of the desired number of bosons just becomes a bound state. And in the case of a fixed interaction one can tune the threshold of the potential such that one remains with the desired number of bosons. With this approach the counting statistics of the problem are fully under control. With just two parameters it is possible to control the interplay of the one-particle potential and the interparticle interactions in order to manufacture any desired final state of $N_{IN}$ ``IN'' bosons and $N_{OUT}$ ``OUT'' bosons. 

The patterns found for the dynamics of the correlations and coherence in momentum space, as depicted in Figs.~\ref{g1k_vs_T},\ref{g2k_vs_T},\ref{corr1_T_N3},\ref{corr2_T_N3} and analyzed in Sections~\ref{cont2} and \ref{cont3} hold also for the present case of $N=101$ bosons. Since the first and second order momentum correlation functions $g^{(1)}$ and $g^{(2)}$ are similar to the ones in Figs.~\ref{g1k_vs_T},\ref{g2k_vs_T},\ref{corr1_T_N3},\ref{corr2_T_N3} they are not shown here for the sake of brevity. It is worthwhile to state here, that the distance between the lines decreases when the particle number is increasing and the minima between the peaks become less pronounced. Hence, the lines are not as clearly visible in the case of a large particle number. The momenta at which first and second order coherence is lost are predicted by the model consideration, i.e., Eq.~\eqref{k_threshold} and Section~\ref{modelv2}. To predict these momenta for which the coherence is lost one has to find the velocities at 
which particles are escaping for given $T$ and $\lambda_0$. To achieve this, one first determines the number of atoms which is still bound for the chosen $T$ and $\lambda_0$. Subsequently, Eq.~\eqref{k_threshold} is used to determine the momenta of the escaping particles. These momenta are the momenta where the coherence is lost. Furthermore, when one adjusts $\lambda_0$,$N$ and $T$ a many-body wavefunction which shows first and second order incoherence for a certain set of momenta can be manufactured -- such a control on the coherence properties of the wavefunction of a many-boson system could be very useful to study the coherence in atom-laser experiments~\cite{BECketterle,atoml1,atoml2,atoml3}.
\vspace*{-3mm}

\section{Brief Summary and Outlook}\label{fin}
In summary, this paper has shown that one can fully control the number of ejected particles and their momentum density in the tunneling to open space process. Namely, the final state's counting statistics can be managed at will, i.e., how many particles will reside in the interior ``IN'' and exterior ``OUT'' regions after the process completes. This can be achieved by manipulating the interplay of the threshold of the potential and the two-body interaction. The overall many-boson process is made up of single-particle processes which are well-described using the following model (cf. Fig.~\ref{model_mu_T}): The momentum $k_i^T$ of the ejection process is defined by the chemical potential $\mu_i$ of the confined $N_{IN}$-boson system. This chemical potential $\mu_i$ is firstly used to overcome the potential threshold $T$ and the remainder of the energy is subsequently converted to kinetic energy $E_{kin}(T,\mu_i)$. Hence, one finds peaks in the momentum distributions at $k_i^T=\sqrt{2m (\mu_i -T)}$. At 
precisely the peaks' positions, the first-oder coherence of the bosons is lost, but the second order coherence is almost completely preserved. This shows the one-particle nature of the processes. Employing this characteristic, one can control the structure of the coherence and correlations in the process by switching on or off certain processes with the interaction $\lambda_0$ or the threshold $T$. 
The model (cf. Fig.~\ref{model_mu_T} and Section~\ref{modelv2}) is a consideration on the energies of $N_{IN}$ confined bosons, $E_{IN}$, and $N_{OUT}$ escaped bosons in open space, $E_{OUT}$. In the present case, $E_{IN}$ is the energy of $N_{IN}$ parabolically trapped interacting particles and $E_{OUT}$ is defined by the threshold $T$ and the difference in $E_{IN}$ for varying $N_{IN}$. In this respect, the control exerted by the potential threshold $T$ and the interaction $\lambda_0$ means to adjust $E_{OUT}$ and $E_{IN}$, respectively. Since the external potential and the interparticle interactions can be controlled almost at will, the system's dynamics are also under full control. This renders it an good candidate for a quantum simulator of other processes with similar properties. Such processes include complicated multiple ionization or dissociation processes \cite{photoass_tun,distun}, where one could use the threshold to adjust the peaks in the momentum distribution such that they correspond to the 
ionization energies and the interaction to tune the peaks in the momentum distribution to mimick the ionization spectrum. Furthermore, the control schemes and found physics of the dynamics especially in the exterior ``OUT'' part of the potential are very similar to atom lasers \cite{BECketterle,atoml1,atoml2,atoml3} and could be useful to simulate and design their physical properties. \vspace*{-2cm}
 
\cleardoublepage

\begin{acknowledgements}
Financial support by the Deutsche Forschungsgemeinschaft, the Minerva Foundation, the HGS MathComp is gratefully acknowledged. Computation time on the Cray XE6 system Hermit and the NEC Nehalem cluster Laki at the H\"ochstleitsungsrechenzentrum Stuttgart and the bwGRiD-initiative in Esslingen, Ulm, Mannheim, Freiburg, and Stuttgart are gratefully acknowledged.
\end{acknowledgements}

\appendix

\section{The multiconfigurational time-dependent Hartree method for bosons and computational details}\label{MCTDHB}

The details and derivation of the used computational method, MCTDHB, are given in Ref.~\cite{alon:08}. MCTDHB is capable of providing numerically exact solutions of the time-dependent many-boson Schr\"odinger equation (TDSE), see Ref.~\cite{axel_exact}. The method relies on expanding the wavefunction with multiple, time-dependent configurations $\vert \vec{n};t \rangle$ weighted with time-dependent coefficients $C_{\vec{n}}(t)$:

\begin{equation}
 \vert \Psi \rangle =\sum_{\vec{n}} C_{\vec{n}} (t) \vert \vec{n}; t \rangle.
\end{equation}

The configurations are a many-body basis build by applying creation and annihilation operators in at most $M$ single-particle time-adaptive states to the quantum mechanical vacuum $\vert vac \rangle$:
\begin{equation}
\vert \vec{n};t \rangle = \vert n_1,n_2,...,n_M; t \rangle= \frac{1}{\sqrt{\prod_{i=1}^M n_i!}} [\hat{b}_1^\dagger(t)]^{n_1} [\hat{b}_2^\dagger(t)]^{n_2} \cdots [\hat{b}_M^\dagger(t)]^{n_M}  \vert vac \rangle.
\end{equation}
The MCTDHB equations of motion are obtained by tackling the TDSE with the time-dependent variational principle and requiring the stationarity of the resulting functional action when varying the coefficients $C_{\vec{n}}(t)$ and the single-particle states $\hat{b}_k(t),\hat{b}_k^\dagger(t)$, see Ref.~\cite{alon:08}. The $\binom{N+M-1}{N}$ linear equations of motion for the coefficients are coupled to the $M$ non-linear integrodifferential equations of motion of the orbitals. Since the derivation is variational and the basis used is a formally complete set in the limit $M\rightarrow\infty$, convergence with respect to the number of orbitals implies the convergence to the exact solution of TDSE for the problem under consideration \cite{axel_exact}. The use of time-adaptive orbitals is of key importance for the achievement of numerical exactness: a much smaller number of time-adaptive orbitals is needed to achieve the same level of accuracy as compared to the number of basis functions in a static, time-
independent basis \cite{axel_exact}.

In the present case the MCTDHB software package \cite{Streltsov2010} was employed to obtain such converged solutions of the TDSE. The computations used grids of sizes of up to $[-5;7465]$ in dimensionless units, sampled by up to $2^{16}=65536$ time-independent basis functions (grid points) with up to $M=14$ orbitals.

\section{Polynomial connection of the parabolic potential and its threshold}\label{polynomial}
There are four constraints to the polynomial continuation, namely that both the polynomial itself and its first derivatives have to be equal to the values and first derivatives of the neighboring potential at $x_{c1}=2$ and $x_{c2}=4$, see Fig.~\ref{pot-tnot0}. Therefore, a polynomial of at least third order with four coefficients, $A,B,C,D$, is required:
\begin{equation}
P(x)=Ax^3+Bx^2+Cx+D.\label{poly}
\end{equation}
With the constraints
\begin{eqnarray}
P(x_{c1})=Ax_{c1}^3+Bx_{c1}^2+Cx_{c1}+D=V_h(x_{c1})=2, \label{polya}\\
\frac{d}{dx}P(x)\mid_{x=x_{c1}}=3Ax_{c1}^2+2Bx_{c1}+C=\frac{d}{dx}V_h(x)\mid_{x=x_{c1}}=2\label{polyb}
\end{eqnarray}
for the connection at $x_{c1}$ to the harmonic trapping potential $V_h(x)$ and
\begin{eqnarray}
P(x_{c2})=Ax_{c2}^3+Bx_{c2}^2+Cx_{c2}+D=T,\label{polyc} \\
\frac{d}{dx}P(x)\mid_{x=x_{c2}}=3Ax_{c2}^2+2Bx_{c2}+C=\frac{d}{dx}T=0\label{polyd}
\end{eqnarray}
for the connection to the constant threshold $T$ at $x_{c2}$. From these four equations the coefficients $A(T),B(T),C(T),D(T)$ can be obtained easily, cf. Table~\ref{tab1}. One can hence control the threshold $T$ arbitrarily while maintaining a smooth potential.
The overall potential part of the Hamiltonian then reads:
\begin{equation}
V(x)=\Theta(x_{c1}-x)\cdot \frac{1}{2}x^2 + \Theta(x-x_{c1}) \cdot \Theta(x_{c2}-x) \cdot P(x) + \Theta(x-x_{c2})\cdot T
\end{equation}
Here $\Theta(\cdot)$ is the Heaviside step function.

\begin{table}[!]
\begin{center}
\begin{tabular}{|l|r|}
\hline
Coefficient & Value at T=0.5\\ \hline
$A(T)=-\frac{1}{4}T+1$           &  $0.875$        \\ \hline
$B(T)=\frac{9}{4}T-\frac{19}{2}$ &  $8.375$        \\ \hline
$C(T)=-6T+28 $                   &  $25.0$         \\ \hline
$D(T)=5T-24$                     &  $21.5$         \\ \hline
\end{tabular}
\caption{Parameters of the potential with a threshold and their dependence on the threshold, cf. Eqs.~\eqref{poly},\eqref{polya},\eqref{polyb},\eqref{polyc},\eqref{polyd}.}\label{tab1}
\end{center}
\end{table}

\clearpage
\bibliographystyle{apsrev}
\bibliography{t_and_t}

\begin{thebibliography}{59}
\expandafter\ifx\csname natexlab\endcsname\relax\def\natexlab#1{#1}\fi
\expandafter\ifx\csname bibnamefont\endcsname\relax
  \def\bibnamefont#1{#1}\fi
\expandafter\ifx\csname bibfnamefont\endcsname\relax
  \def\bibfnamefont#1{#1}\fi
\expandafter\ifx\csname citenamefont\endcsname\relax
  \def\citenamefont#1{#1}\fi
\expandafter\ifx\csname url\endcsname\relax
  \def\url#1{\texttt{#1}}\fi
\expandafter\ifx\csname urlprefix\endcsname\relax\def\urlprefix{URL }\fi
\providecommand{\bibinfo}[2]{#2}
\providecommand{\eprint}[2][]{\url{#2}}

\bibitem[{\citenamefont{Morsch and Oberthaler}(2006)}]{oberthaler}
\bibinfo{author}{\bibfnamefont{O.}~\bibnamefont{Morsch}} \bibnamefont{and}
  \bibinfo{author}{\bibfnamefont{M.~K.} \bibnamefont{Oberthaler}},
  \bibinfo{journal}{Rev. Mod. Phys.} \textbf{\bibinfo{volume}{\textbf{78}}},
  \bibinfo{pages}{179} (\bibinfo{year}{2006}).

\bibitem[{\citenamefont{Greiner et~al.}(2002)\citenamefont{Greiner, Mandel,
  Esslinger, H\"ansch, and Bloch}}]{greiner:02}
\bibinfo{author}{\bibfnamefont{M.}~\bibnamefont{Greiner}},
  \bibinfo{author}{\bibfnamefont{O.}~\bibnamefont{Mandel}},
  \bibinfo{author}{\bibfnamefont{T.}~\bibnamefont{Esslinger}},
  \bibinfo{author}{\bibfnamefont{T.~W.} \bibnamefont{H\"ansch}},
  \bibnamefont{and} \bibinfo{author}{\bibfnamefont{I.}~\bibnamefont{Bloch}},
  \bibinfo{journal}{Nature (London)} \textbf{\bibinfo{volume}{415}},
  \bibinfo{pages}{39} (\bibinfo{year}{2002}).

\bibitem[{\citenamefont{Ketterle}(2002)}]{BECketterle}
\bibinfo{author}{\bibfnamefont{W.}~\bibnamefont{Ketterle}},
  \bibinfo{journal}{Rev. Mod. Phys.} \textbf{\bibinfo{volume}{\textbf{74}}},
  \bibinfo{pages}{1131} (\bibinfo{year}{2002}).

\bibitem[{\citenamefont{Anderson et~al.}(1995)\citenamefont{Anderson, Ensher,
  Matthews, Wiemann, and Cornell}}]{anderson:95}
\bibinfo{author}{\bibfnamefont{M.~H.} \bibnamefont{Anderson}},
  \bibinfo{author}{\bibfnamefont{J.~R.} \bibnamefont{Ensher}},
  \bibinfo{author}{\bibfnamefont{M.~R.} \bibnamefont{Matthews}},
  \bibinfo{author}{\bibfnamefont{C.~E.} \bibnamefont{Wiemann}},
  \bibnamefont{and} \bibinfo{author}{\bibfnamefont{E.~A.}
  \bibnamefont{Cornell}}, \bibinfo{journal}{Science}
  \textbf{\bibinfo{volume}{269}}, \bibinfo{pages}{198} (\bibinfo{year}{1995}).

\bibitem[{\citenamefont{Bradley et~al.}(1995)\citenamefont{Bradley, Sackett,
  Tollet, and Hulet}}]{bradley:95}
\bibinfo{author}{\bibfnamefont{C.~C.} \bibnamefont{Bradley}},
  \bibinfo{author}{\bibfnamefont{C.~A.} \bibnamefont{Sackett}},
  \bibinfo{author}{\bibfnamefont{J.~J.} \bibnamefont{Tollet}},
  \bibnamefont{and} \bibinfo{author}{\bibfnamefont{R.~G.} \bibnamefont{Hulet}},
  \bibinfo{journal}{Phys. Rev. Lett.} \textbf{\bibinfo{volume}{75}},
  \bibinfo{pages}{1687} (\bibinfo{year}{1995}).

\bibitem[{\citenamefont{Henderson et~al.}(2009)\citenamefont{Henderson, Ryu,
  MacCormick, and Boshier}}]{Henderson:09}
\bibinfo{author}{\bibfnamefont{K.}~\bibnamefont{Henderson}},
  \bibinfo{author}{\bibfnamefont{C.}~\bibnamefont{Ryu}},
  \bibinfo{author}{\bibfnamefont{C.}~\bibnamefont{MacCormick}},
  \bibnamefont{and} \bibinfo{author}{\bibfnamefont{M.~G.}
  \bibnamefont{Boshier}}, \bibinfo{journal}{New J. Phys.}
  \textbf{\bibinfo{volume}{11}}, \bibinfo{pages}{043030}
  (\bibinfo{year}{2009}).

\bibitem[{\citenamefont{Chin et~al.}(2010)\citenamefont{Chin, Grimm, Julienne,
  and Tiesinga}}]{chin:10}
\bibinfo{author}{\bibfnamefont{C.}~\bibnamefont{Chin}},
  \bibinfo{author}{\bibfnamefont{R.}~\bibnamefont{Grimm}},
  \bibinfo{author}{\bibfnamefont{P.}~\bibnamefont{Julienne}}, \bibnamefont{and}
  \bibinfo{author}{\bibfnamefont{E.}~\bibnamefont{Tiesinga}},
  \bibinfo{journal}{Rev. Mod. Phys.} \textbf{\bibinfo{volume}{82}},
  \bibinfo{pages}{1225} (\bibinfo{year}{2010}).

\bibitem[{\citenamefont{Görlitz et~al.}(2001)\citenamefont{Görlitz, Vogels,
  Leanhardt, Raman, Gustavson, Abo-Shaeer, Chikkatur, Gupta, Inouye, Rosenband
  et~al.}}]{1d1}
\bibinfo{author}{\bibfnamefont{A.}~\bibnamefont{Görlitz}},
  \bibinfo{author}{\bibfnamefont{J.~M.} \bibnamefont{Vogels}},
  \bibinfo{author}{\bibfnamefont{A.~E.} \bibnamefont{Leanhardt}},
  \bibinfo{author}{\bibfnamefont{C.}~\bibnamefont{Raman}},
  \bibinfo{author}{\bibfnamefont{T.~L.} \bibnamefont{Gustavson}},
  \bibinfo{author}{\bibfnamefont{J.~R.} \bibnamefont{Abo-Shaeer}},
  \bibinfo{author}{\bibfnamefont{A.~P.} \bibnamefont{Chikkatur}},
  \bibinfo{author}{\bibfnamefont{S.}~\bibnamefont{Gupta}},
  \bibinfo{author}{\bibfnamefont{S.}~\bibnamefont{Inouye}},
  \bibinfo{author}{\bibfnamefont{T.}~\bibnamefont{Rosenband}},
  \bibnamefont{et~al.}, \bibinfo{journal}{Phys. Rev. Lett.}
  \textbf{\bibinfo{volume}{\textbf{87}}}, \bibinfo{pages}{130402}
  (\bibinfo{year}{2001}).

\bibitem[{\citenamefont{Schreck et~al.}(2001)\citenamefont{Schreck, Khaykovich,
  Corwin, Ferrari, Bourdel, Cubizolles, and Salomon}}]{1d2}
\bibinfo{author}{\bibfnamefont{F.}~\bibnamefont{Schreck}},
  \bibinfo{author}{\bibfnamefont{L.}~\bibnamefont{Khaykovich}},
  \bibinfo{author}{\bibfnamefont{K.~L.} \bibnamefont{Corwin}},
  \bibinfo{author}{\bibfnamefont{G.}~\bibnamefont{Ferrari}},
  \bibinfo{author}{\bibfnamefont{T.}~\bibnamefont{Bourdel}},
  \bibinfo{author}{\bibfnamefont{J.}~\bibnamefont{Cubizolles}},
  \bibnamefont{and} \bibinfo{author}{\bibfnamefont{C.}~\bibnamefont{Salomon}},
  \bibinfo{journal}{Phys. Rev. Lett.} \textbf{\bibinfo{volume}{\textbf{87}}},
  \bibinfo{pages}{080403} (\bibinfo{year}{2001}).

\bibitem[{\citenamefont{Greiner et~al.}(2001)\citenamefont{Greiner, Bloch,
  Mandel, Hänsch, and Esslinger}}]{1d3}
\bibinfo{author}{\bibfnamefont{M.}~\bibnamefont{Greiner}},
  \bibinfo{author}{\bibfnamefont{I.}~\bibnamefont{Bloch}},
  \bibinfo{author}{\bibfnamefont{O.}~\bibnamefont{Mandel}},
  \bibinfo{author}{\bibfnamefont{T.~W.} \bibnamefont{Hänsch}},
  \bibnamefont{and}
  \bibinfo{author}{\bibfnamefont{T.}~\bibnamefont{Esslinger}},
  \bibinfo{journal}{Phys. Rev. Lett.} \textbf{\bibinfo{volume}{\textbf{87}}},
  \bibinfo{pages}{160405} (\bibinfo{year}{2001}).

\bibitem[{\citenamefont{Bloch et~al.}(2008)\citenamefont{Bloch, Dalibard, and
  Zwerger}}]{bloch:08}
\bibinfo{author}{\bibfnamefont{I.}~\bibnamefont{Bloch}},
  \bibinfo{author}{\bibfnamefont{J.}~\bibnamefont{Dalibard}}, \bibnamefont{and}
  \bibinfo{author}{\bibfnamefont{W.}~\bibnamefont{Zwerger}},
  \bibinfo{journal}{Rev. Mod. Phys.} \textbf{\bibinfo{volume}{80}},
  \bibinfo{pages}{885} (\bibinfo{year}{2008}).

\bibitem[{\citenamefont{Hemmerich and Smith}(2007)}]{OL_solid_state1}
\bibinfo{author}{\bibfnamefont{A.}~\bibnamefont{Hemmerich}} \bibnamefont{and}
  \bibinfo{author}{\bibfnamefont{C.~M.} \bibnamefont{Smith}},
  \bibinfo{journal}{Phys. Rev. Lett.} \textbf{\bibinfo{volume}{99}},
  \bibinfo{pages}{113002} (\bibinfo{year}{2007}).

\bibitem[{\citenamefont{Pu et~al.}(2001)\citenamefont{Pu, Zhang, and
  Meystre}}]{OL_solid_state2}
\bibinfo{author}{\bibfnamefont{H.}~\bibnamefont{Pu}},
  \bibinfo{author}{\bibfnamefont{W.}~\bibnamefont{Zhang}}, \bibnamefont{and}
  \bibinfo{author}{\bibfnamefont{P.}~\bibnamefont{Meystre}},
  \bibinfo{journal}{Phys. Rev. Lett.} \textbf{\bibinfo{volume}{87}},
  \bibinfo{pages}{140405} (\bibinfo{year}{2001}).

\bibitem[{\citenamefont{Lahav et~al.}(2010)\citenamefont{Lahav, Itah, Blumkin,
  Gordon, Rinott, Zayats, and Steinhauer}}]{steinhauer:10}
\bibinfo{author}{\bibfnamefont{O.}~\bibnamefont{Lahav}},
  \bibinfo{author}{\bibfnamefont{A.}~\bibnamefont{Itah}},
  \bibinfo{author}{\bibfnamefont{A.}~\bibnamefont{Blumkin}},
  \bibinfo{author}{\bibfnamefont{C.}~\bibnamefont{Gordon}},
  \bibinfo{author}{\bibfnamefont{S.}~\bibnamefont{Rinott}},
  \bibinfo{author}{\bibfnamefont{A.}~\bibnamefont{Zayats}}, \bibnamefont{and}
  \bibinfo{author}{\bibfnamefont{J.}~\bibnamefont{Steinhauer}},
  \bibinfo{journal}{Phys. Rev. Lett.} \textbf{\bibinfo{volume}{105}},
  \bibinfo{pages}{240401} (\bibinfo{year}{2010}).

\bibitem[{\citenamefont{Macher and Parentani}(2009)}]{macher:09}
\bibinfo{author}{\bibfnamefont{J.}~\bibnamefont{Macher}} \bibnamefont{and}
  \bibinfo{author}{\bibfnamefont{R.}~\bibnamefont{Parentani}},
  \bibinfo{journal}{Phys. Rev. A} \textbf{\bibinfo{volume}{{\bf 80}}},
  \bibinfo{pages}{043601} (\bibinfo{year}{2009}).

\bibitem[{\citenamefont{Jaskula et~al.}(2012)\citenamefont{Jaskula, Partridge,
  Bonneau, Lopes, Ruaudel, Boiron, and Westbrook}}]{westbrook:12}
\bibinfo{author}{\bibfnamefont{J.-C.} \bibnamefont{Jaskula}},
  \bibinfo{author}{\bibfnamefont{G.~B.} \bibnamefont{Partridge}},
  \bibinfo{author}{\bibfnamefont{M.}~\bibnamefont{Bonneau}},
  \bibinfo{author}{\bibfnamefont{R.}~\bibnamefont{Lopes}},
  \bibinfo{author}{\bibfnamefont{J.}~\bibnamefont{Ruaudel}},
  \bibinfo{author}{\bibfnamefont{D.}~\bibnamefont{Boiron}}, \bibnamefont{and}
  \bibinfo{author}{\bibfnamefont{C.~I.} \bibnamefont{Westbrook}},
  \bibinfo{journal}{Phys. Rev. Lett.} \textbf{\bibinfo{volume}{109}},
  \bibinfo{pages}{220401} (\bibinfo{year}{2012}).

\bibitem[{\citenamefont{Razavy}(2003)}]{tun_book}
\bibinfo{author}{\bibfnamefont{M.}~\bibnamefont{Razavy}},
  \emph{\bibinfo{title}{Quantum Theory of Tunneling}}
  (\bibinfo{publisher}{World Scientific Publishing Co., Singapore},
  \bibinfo{year}{2003}).

\bibitem[{\citenamefont{Kramers}(1926)}]{WKB}
\bibinfo{author}{\bibfnamefont{H.~A.} \bibnamefont{Kramers}},
  \bibinfo{journal}{Zeitschr. f. Physik A} \textbf{\bibinfo{volume}{39}},
  \bibinfo{pages}{828} (\bibinfo{year}{1926}).

\bibitem[{\citenamefont{Gurney and Condon}(1928)}]{tunstart1}
\bibinfo{author}{\bibfnamefont{R.~W.} \bibnamefont{Gurney}} \bibnamefont{and}
  \bibinfo{author}{\bibfnamefont{E.~U.} \bibnamefont{Condon}},
  \bibinfo{journal}{Nature (London)} \textbf{\bibinfo{volume}{122}},
  \bibinfo{pages}{439} (\bibinfo{year}{1928}).

\bibitem[{\citenamefont{Gurney and Condon}(1929)}]{tunstart2}
\bibinfo{author}{\bibfnamefont{R.~W.} \bibnamefont{Gurney}} \bibnamefont{and}
  \bibinfo{author}{\bibfnamefont{E.~U.} \bibnamefont{Condon}},
  \bibinfo{journal}{Phys. Rev.} \textbf{\bibinfo{volume}{33}},
  \bibinfo{pages}{127} (\bibinfo{year}{1929}).

\bibitem[{\citenamefont{Kim and Brand}(2011)}]{Brand:11}
\bibinfo{author}{\bibfnamefont{S.}~\bibnamefont{Kim}} \bibnamefont{and}
  \bibinfo{author}{\bibfnamefont{J.}~\bibnamefont{Brand}}, \bibinfo{journal}{J.
  Phys. B: At. Mol. Opt. Phys.} \textbf{\bibinfo{volume}{44}},
  \bibinfo{pages}{195301} (\bibinfo{year}{2011}).

\bibitem[{\citenamefont{{Glick} and {Carr}}(2011)}]{LINCOLN}
\bibinfo{author}{\bibfnamefont{J.~A.} \bibnamefont{{Glick}}} \bibnamefont{and}
  \bibinfo{author}{\bibfnamefont{L.~D.} \bibnamefont{{Carr}}},
  \bibinfo{journal}{ArXiv}  (\bibinfo{year}{2011}), \eprint{1105.5164}.

\bibitem[{\citenamefont{del Campo et~al.}(2006)\citenamefont{del Campo,
  Delgado, Garc\'{\i}a-Calder\'{o}n, Muga, and Raizen}}]{TGtunneling}
\bibinfo{author}{\bibfnamefont{A.}~\bibnamefont{del Campo}},
  \bibinfo{author}{\bibfnamefont{F.}~\bibnamefont{Delgado}},
  \bibinfo{author}{\bibfnamefont{G.}~\bibnamefont{Garc\'{\i}a-Calder\'{o}n}},
  \bibinfo{author}{\bibfnamefont{J.}~\bibnamefont{Muga}}, \bibnamefont{and}
  \bibinfo{author}{\bibfnamefont{M.}~\bibnamefont{Raizen}},
  \bibinfo{journal}{Phys. Rev. A} \textbf{\bibinfo{volume}{\textbf{74}}}
  (\bibinfo{year}{2006}).

\bibitem[{\citenamefont{del Campo and Muga}(2006)}]{muga2}
\bibinfo{author}{\bibfnamefont{A.}~\bibnamefont{del Campo}} \bibnamefont{and}
  \bibinfo{author}{\bibfnamefont{J.}~\bibnamefont{Muga}},
  \bibinfo{journal}{Europhys. Lett.} \textbf{\bibinfo{volume}{\textbf{6}}},
  \bibinfo{pages}{965} (\bibinfo{year}{2006}).

\bibitem[{\citenamefont{Hunn et~al.}(2013)\citenamefont{Hunn, Zimmermann,
  Hiller, and Buchleitner}}]{PhysRevA.87.043626}
\bibinfo{author}{\bibfnamefont{S.}~\bibnamefont{Hunn}},
  \bibinfo{author}{\bibfnamefont{K.}~\bibnamefont{Zimmermann}},
  \bibinfo{author}{\bibfnamefont{M.}~\bibnamefont{Hiller}}, \bibnamefont{and}
  \bibinfo{author}{\bibfnamefont{A.}~\bibnamefont{Buchleitner}},
  \bibinfo{journal}{Phys. Rev. A} \textbf{\bibinfo{volume}{87}},
  \bibinfo{pages}{043626} (\bibinfo{year}{2013}).

\bibitem[{\citenamefont{Carr et~al.}(2005)\citenamefont{Carr, Holland, and
  Malomed}}]{MQTBECWKB}
\bibinfo{author}{\bibfnamefont{L.}~\bibnamefont{Carr}},
  \bibinfo{author}{\bibfnamefont{M.}~\bibnamefont{Holland}}, \bibnamefont{and}
  \bibinfo{author}{\bibfnamefont{B.}~\bibnamefont{Malomed}},
  \bibinfo{journal}{Journal Phys. B} \textbf{\bibinfo{volume}{\textbf{38}}},
  \bibinfo{pages}{3217} (\bibinfo{year}{2005}).

\bibitem[{\citenamefont{Moiseyev and Cederbaum}(2005)}]{lenztunfrag}
\bibinfo{author}{\bibfnamefont{N.}~\bibnamefont{Moiseyev}} \bibnamefont{and}
  \bibinfo{author}{\bibfnamefont{L.}~\bibnamefont{Cederbaum}},
  \bibinfo{journal}{Phys. Rev. A} \textbf{\bibinfo{volume}{\textbf{72}}}
  (\bibinfo{year}{2005}).

\bibitem[{\citenamefont{Schlagheck and Wimberger}(2007)}]{nonexpCS}
\bibinfo{author}{\bibfnamefont{P.}~\bibnamefont{Schlagheck}} \bibnamefont{and}
  \bibinfo{author}{\bibfnamefont{S.}~\bibnamefont{Wimberger}},
  \bibinfo{journal}{Applied Phys. B} \textbf{\bibinfo{volume}{\textbf{86}}},
  \bibinfo{pages}{385} (\bibinfo{year}{2007}).

\bibitem[{\citenamefont{Lode et~al.}(2009)\citenamefont{Lode, Streltsov, Alon,
  Meyer, and Cederbaum}}]{axel:09}
\bibinfo{author}{\bibfnamefont{A.~U.~J.} \bibnamefont{Lode}},
  \bibinfo{author}{\bibfnamefont{A.~I.} \bibnamefont{Streltsov}},
  \bibinfo{author}{\bibfnamefont{O.~E.} \bibnamefont{Alon}},
  \bibinfo{author}{\bibfnamefont{H.-D.} \bibnamefont{Meyer}}, \bibnamefont{and}
  \bibinfo{author}{\bibfnamefont{L.~S.} \bibnamefont{Cederbaum}},
  \bibinfo{journal}{J. Phys. B} \textbf{\bibinfo{volume}{42}},
  \bibinfo{pages}{044018} (\bibinfo{year}{2009}).

\bibitem[{\citenamefont{Lode et~al.}(2010)\citenamefont{Lode, Streltsov, Alon,
  Meyer, and Cederbaum}}]{axel:10}
\bibinfo{author}{\bibfnamefont{A.~U.~J.} \bibnamefont{Lode}},
  \bibinfo{author}{\bibfnamefont{A.~I.} \bibnamefont{Streltsov}},
  \bibinfo{author}{\bibfnamefont{O.~E.} \bibnamefont{Alon}},
  \bibinfo{author}{\bibfnamefont{H.-D.} \bibnamefont{Meyer}}, \bibnamefont{and}
  \bibinfo{author}{\bibfnamefont{L.~S.} \bibnamefont{Cederbaum}},
  \bibinfo{journal}{J. Phys. B} \textbf{\bibinfo{volume}{43}},
  \bibinfo{pages}{029802} (\bibinfo{year}{2010}).

\bibitem[{\citenamefont{{Lode} et~al.}(2012)\citenamefont{{Lode}, {Streltsov},
  {Sakmann}, {Alon}, and {Cederbaum}}}]{axel:12}
\bibinfo{author}{\bibfnamefont{A.~U.~J.} \bibnamefont{{Lode}}},
  \bibinfo{author}{\bibfnamefont{A.~I.} \bibnamefont{{Streltsov}}},
  \bibinfo{author}{\bibfnamefont{K.}~\bibnamefont{{Sakmann}}},
  \bibinfo{author}{\bibfnamefont{O.~E.} \bibnamefont{{Alon}}},
  \bibnamefont{and} \bibinfo{author}{\bibfnamefont{L.~S.}
  \bibnamefont{{Cederbaum}}}, \bibinfo{journal}{Proc. Natl. Acad. Sci. USA}
  \textbf{\bibinfo{volume}{109}}, \bibinfo{pages}{13521}
  (\bibinfo{year}{2012}).

\bibitem[{\citenamefont{Nozi{\`e}res and James}(1982)}]{nozieres:82}
\bibinfo{author}{\bibfnamefont{P.}~\bibnamefont{Nozi{\`e}res}}
  \bibnamefont{and} \bibinfo{author}{\bibfnamefont{D.~S.} \bibnamefont{James}},
  \bibinfo{journal}{J. Phys. (Paris)} \textbf{\bibinfo{volume}{43}},
  \bibinfo{pages}{1133} (\bibinfo{year}{1982}).

\bibitem[{\citenamefont{Nozi{\`e}res}(1996)}]{nozieres:96}
\bibinfo{author}{\bibfnamefont{P.}~\bibnamefont{Nozi{\`e}res}},
  \emph{\bibinfo{title}{Bose-Einstein Condensation}}
  (\bibinfo{publisher}{Cambridge University Press, New York},
  \bibinfo{year}{1996}), pp. \bibinfo{pages}{15--30}.

\bibitem[{\citenamefont{Spekkens and Sipe}(1999)}]{spekkens:99}
\bibinfo{author}{\bibfnamefont{R.~W.} \bibnamefont{Spekkens}} \bibnamefont{and}
  \bibinfo{author}{\bibfnamefont{J.~E.} \bibnamefont{Sipe}},
  \bibinfo{journal}{Phys. Rev. A} \textbf{\bibinfo{volume}{59}},
  \bibinfo{pages}{3868} (\bibinfo{year}{1999}).

\bibitem[{\citenamefont{Alon and Cederbaum}(2005)}]{alon.prl2:05}
\bibinfo{author}{\bibfnamefont{O.~E.} \bibnamefont{Alon}} \bibnamefont{and}
  \bibinfo{author}{\bibfnamefont{L.~S.} \bibnamefont{Cederbaum}},
  \bibinfo{journal}{Phys. Rev. Lett.} \textbf{\bibinfo{volume}{95}},
  \bibinfo{pages}{140402} (\bibinfo{year}{2005}).

\bibitem[{\citenamefont{Klaiman et~al.}(2006)\citenamefont{Klaiman, Moiseyev,
  and Cederbaum}}]{lenzexact2}
\bibinfo{author}{\bibfnamefont{S.}~\bibnamefont{Klaiman}},
  \bibinfo{author}{\bibfnamefont{N.}~\bibnamefont{Moiseyev}}, \bibnamefont{and}
  \bibinfo{author}{\bibfnamefont{L.}~\bibnamefont{Cederbaum}},
  \bibinfo{journal}{Phys. Rev. A} \textbf{\bibinfo{volume}{\textbf{73}}},
  \bibinfo{pages}{013622} (\bibinfo{year}{2006}).

\bibitem[{\citenamefont{M\"{u}ller et~al.}(2006)\citenamefont{M\"{u}ller, Ho,
  Ueda, and Baym}}]{fragmentothers2}
\bibinfo{author}{\bibfnamefont{E.}~\bibnamefont{M\"{u}ller}},
  \bibinfo{author}{\bibfnamefont{T.-L.} \bibnamefont{Ho}},
  \bibinfo{author}{\bibfnamefont{M.}~\bibnamefont{Ueda}}, \bibnamefont{and}
  \bibinfo{author}{\bibfnamefont{G.}~\bibnamefont{Baym}},
  \bibinfo{journal}{Phys. Rev. A} \textbf{\bibinfo{volume}{\textbf{74}}},
  \bibinfo{pages}{033612} (\bibinfo{year}{2006}).

\bibitem[{\citenamefont{Streltsov et~al.}(2006)\citenamefont{Streltsov, Alon,
  and Cederbaum}}]{streltsov:06}
\bibinfo{author}{\bibfnamefont{A.~I.} \bibnamefont{Streltsov}},
  \bibinfo{author}{\bibfnamefont{O.~E.} \bibnamefont{Alon}}, \bibnamefont{and}
  \bibinfo{author}{\bibfnamefont{L.~S.} \bibnamefont{Cederbaum}},
  \bibinfo{journal}{Phys. Rev. A} \textbf{\bibinfo{volume}{73}},
  \bibinfo{pages}{063626} (\bibinfo{year}{2006}).

\bibitem[{\citenamefont{Titulaer and Glauber}(1965)}]{Glauber}
\bibinfo{author}{\bibfnamefont{U.}~\bibnamefont{Titulaer}} \bibnamefont{and}
  \bibinfo{author}{\bibfnamefont{R.}~\bibnamefont{Glauber}},
  \bibinfo{journal}{Phys. Rev.} \textbf{\bibinfo{volume}{\textbf{140}}},
  \bibinfo{pages}{B676} (\bibinfo{year}{1965}).

\bibitem[{\citenamefont{Glauber}(1963)}]{Glauber1}
\bibinfo{author}{\bibfnamefont{R.~J.} \bibnamefont{Glauber}},
  \bibinfo{journal}{Phys. Rev.} \textbf{\bibinfo{volume}{\textbf{130}}},
  \bibinfo{pages}{6} (\bibinfo{year}{1963}).

\bibitem[{\citenamefont{Streltsov et~al.}(2007)\citenamefont{Streltsov, Alon,
  and Cederbaum}}]{alexejsplit}
\bibinfo{author}{\bibfnamefont{A.}~\bibnamefont{Streltsov}},
  \bibinfo{author}{\bibfnamefont{O.}~\bibnamefont{Alon}}, \bibnamefont{and}
  \bibinfo{author}{\bibfnamefont{L.}~\bibnamefont{Cederbaum}},
  \bibinfo{journal}{Phys. Rev. Lett.} \textbf{\bibinfo{volume}{\textbf{99}}},
  \bibinfo{pages}{030402} (\bibinfo{year}{2007}).

\bibitem[{\citenamefont{Lode et~al.}(2012)\citenamefont{Lode, Sakmann, Alon,
  Cederbaum, and Streltsov}}]{axel_exact}
\bibinfo{author}{\bibfnamefont{A.~U.~J.} \bibnamefont{Lode}},
  \bibinfo{author}{\bibfnamefont{K.}~\bibnamefont{Sakmann}},
  \bibinfo{author}{\bibfnamefont{O.~E.} \bibnamefont{Alon}},
  \bibinfo{author}{\bibfnamefont{L.~S.} \bibnamefont{Cederbaum}},
  \bibnamefont{and} \bibinfo{author}{\bibfnamefont{A.~I.}
  \bibnamefont{Streltsov}}, \bibinfo{journal}{Phys. Rev. A}
  \textbf{\bibinfo{volume}{86}}, \bibinfo{pages}{063606}
  (\bibinfo{year}{2012}).

\bibitem[{\citenamefont{B\v{r}ezinov\'{a}
  et~al.}(2012)\citenamefont{B\v{r}ezinov\'{a}, Lode, Streltsov, Alon,
  Cederbaum, and Burgd\"{o}rfer}}]{Chaos:12}
\bibinfo{author}{\bibfnamefont{I.}~\bibnamefont{B\v{r}ezinov\'{a}}},
  \bibinfo{author}{\bibfnamefont{A.~U.~J.} \bibnamefont{Lode}},
  \bibinfo{author}{\bibfnamefont{A.~I.} \bibnamefont{Streltsov}},
  \bibinfo{author}{\bibfnamefont{O.~E.} \bibnamefont{Alon}},
  \bibinfo{author}{\bibfnamefont{L.~S.} \bibnamefont{Cederbaum}},
  \bibnamefont{and}
  \bibinfo{author}{\bibfnamefont{J.}~\bibnamefont{Burgd\"{o}rfer}},
  \bibinfo{journal}{Phys. Rev. A} \textbf{\bibinfo{volume}{86}},
  \bibinfo{pages}{013630} (\bibinfo{year}{2012}).

\bibitem[{\citenamefont{Sakmann et~al.}(2009)\citenamefont{Sakmann, Streltsov,
  Alon, and Cederbaum}}]{sakmann:09}
\bibinfo{author}{\bibfnamefont{K.}~\bibnamefont{Sakmann}},
  \bibinfo{author}{\bibfnamefont{A.~I.} \bibnamefont{Streltsov}},
  \bibinfo{author}{\bibfnamefont{O.~E.} \bibnamefont{Alon}}, \bibnamefont{and}
  \bibinfo{author}{\bibfnamefont{L.~S.} \bibnamefont{Cederbaum}},
  \bibinfo{journal}{Phys. Rev. Lett.} \textbf{\bibinfo{volume}{103}},
  \bibinfo{pages}{220601} (\bibinfo{year}{2009}).

\bibitem[{\citenamefont{Sakmann et~al.}(2010)\citenamefont{Sakmann, Streltsov,
  Alon, and Cederbaum}}]{sakmann.pra:10}
\bibinfo{author}{\bibfnamefont{K.}~\bibnamefont{Sakmann}},
  \bibinfo{author}{\bibfnamefont{A.~I.} \bibnamefont{Streltsov}},
  \bibinfo{author}{\bibfnamefont{O.~E.} \bibnamefont{Alon}}, \bibnamefont{and}
  \bibinfo{author}{\bibfnamefont{L.~S.} \bibnamefont{Cederbaum}},
  \bibinfo{journal}{Phys. Rev. A} \textbf{\bibinfo{volume}{82}},
  \bibinfo{pages}{013620} (\bibinfo{year}{2010}).

\bibitem[{\citenamefont{Bloch et~al.}(1999)\citenamefont{Bloch, H\"{a}nsch, and
  Esslinger}}]{atoml1}
\bibinfo{author}{\bibfnamefont{I.}~\bibnamefont{Bloch}},
  \bibinfo{author}{\bibfnamefont{T.}~\bibnamefont{H\"{a}nsch}},
  \bibnamefont{and}
  \bibinfo{author}{\bibfnamefont{T.}~\bibnamefont{Esslinger}},
  \bibinfo{journal}{Phys. Rev. Lett.} \textbf{\bibinfo{volume}{82}},
  \bibinfo{pages}{15} (\bibinfo{year}{1999}).

\bibitem[{\citenamefont{\"{O}ttl et~al.}(2005)\citenamefont{\"{O}ttl, Ritter,
  K\"{o}hl, and Esslinger}}]{atoml2}
\bibinfo{author}{\bibfnamefont{A.}~\bibnamefont{\"{O}ttl}},
  \bibinfo{author}{\bibfnamefont{S.}~\bibnamefont{Ritter}},
  \bibinfo{author}{\bibfnamefont{M.}~\bibnamefont{K\"{o}hl}}, \bibnamefont{and}
  \bibinfo{author}{\bibfnamefont{T.}~\bibnamefont{Esslinger}},
  \bibinfo{journal}{Phys. Rev. Lett.} \textbf{\bibinfo{volume}{95}},
  \bibinfo{pages}{090404} (\bibinfo{year}{2005}).

\bibitem[{\citenamefont{K\"{o}hl et~al.}(2005)\citenamefont{K\"{o}hl, Busch,
  M\o{}lmer, Hänsch, and Esslinger}}]{atoml3}
\bibinfo{author}{\bibfnamefont{M.}~\bibnamefont{K\"{o}hl}},
  \bibinfo{author}{\bibfnamefont{T.}~\bibnamefont{Busch}},
  \bibinfo{author}{\bibfnamefont{K.}~\bibnamefont{M\o{}lmer}},
  \bibinfo{author}{\bibfnamefont{T.}~\bibnamefont{Hänsch}}, \bibnamefont{and}
  \bibinfo{author}{\bibfnamefont{T.}~\bibnamefont{Esslinger}},
  \bibinfo{journal}{Phys. Rev. A} \textbf{\bibinfo{volume}{72}},
  \bibinfo{pages}{063618} (\bibinfo{year}{2005}).

\bibitem[{\citenamefont{Vatasescu}(2000)}]{photoass_tun}
\bibinfo{author}{\bibfnamefont{M.~t.} \bibnamefont{Vatasescu}},
  \bibinfo{journal}{Phys. Rev. A} \textbf{\bibinfo{volume}{61}},
  \bibinfo{pages}{044701} (\bibinfo{year}{2000}).

\bibitem[{\citenamefont{Keller and Weiner}(1984)}]{distun}
\bibinfo{author}{\bibfnamefont{J.}~\bibnamefont{Keller}} \bibnamefont{and}
  \bibinfo{author}{\bibfnamefont{J.}~\bibnamefont{Weiner}},
  \bibinfo{journal}{Phys. Rev. A} \textbf{\bibinfo{volume}{29}},
  \bibinfo{pages}{2943} (\bibinfo{year}{1984}).

\bibitem[{\citenamefont{Naraschewski and Glauber}(1999)}]{corrglauber}
\bibinfo{author}{\bibfnamefont{M.}~\bibnamefont{Naraschewski}}
  \bibnamefont{and} \bibinfo{author}{\bibfnamefont{R.~J.}
  \bibnamefont{Glauber}}, \bibinfo{journal}{Phys. Rev. A}
  \textbf{\bibinfo{volume}{\textbf{59}}}, \bibinfo{pages}{6}
  (\bibinfo{year}{1999}).

\bibitem[{\citenamefont{Coleman and Yukalov}(2000)}]{RDMbook}
\bibinfo{author}{\bibfnamefont{A.}~\bibnamefont{Coleman}} \bibnamefont{and}
  \bibinfo{author}{\bibfnamefont{V.}~\bibnamefont{Yukalov}},
  \emph{\bibinfo{title}{Reduced Density Matrices: Coulson's Challenge}}
  (\bibinfo{publisher}{Springer Heidelberg}, \bibinfo{year}{2000}).

\bibitem[{\citenamefont{Sakmann et~al.}(2008)\citenamefont{Sakmann, Streltsov,
  Alon, and Cederbaum}}]{Kaspar}
\bibinfo{author}{\bibfnamefont{K.}~\bibnamefont{Sakmann}},
  \bibinfo{author}{\bibfnamefont{A.~I.} \bibnamefont{Streltsov}},
  \bibinfo{author}{\bibfnamefont{O.~E.} \bibnamefont{Alon}}, \bibnamefont{and}
  \bibinfo{author}{\bibfnamefont{L.~S.} \bibnamefont{Cederbaum}},
  \bibinfo{journal}{Phys. Rev. A} \textbf{\bibinfo{volume}{\textbf{78}}},
  \bibinfo{pages}{023615} (\bibinfo{year}{2008}).

\bibitem[{\citenamefont{Penrose and Onsager}(1956)}]{penrose:56}
\bibinfo{author}{\bibfnamefont{O.}~\bibnamefont{Penrose}} \bibnamefont{and}
  \bibinfo{author}{\bibfnamefont{L.}~\bibnamefont{Onsager}},
  \bibinfo{journal}{Phys. Rev.} \textbf{\bibinfo{volume}{104}},
  \bibinfo{pages}{576} (\bibinfo{year}{1956}).

\bibitem[{\citenamefont{Brown and Twiss}(1956)}]{HBT1}
\bibinfo{author}{\bibfnamefont{R.~H.} \bibnamefont{Brown}} \bibnamefont{and}
  \bibinfo{author}{\bibfnamefont{R.~Q.} \bibnamefont{Twiss}},
  \bibinfo{journal}{Nature (London)} \textbf{\bibinfo{volume}{178}},
  \bibinfo{pages}{1046 } (\bibinfo{year}{1956}).

\bibitem[{\citenamefont{Brown and Twiss}(1957)}]{HBT2}
\bibinfo{author}{\bibfnamefont{R.~H.} \bibnamefont{Brown}} \bibnamefont{and}
  \bibinfo{author}{\bibfnamefont{R.~Q.} \bibnamefont{Twiss}},
  \bibinfo{journal}{Proc. R. Soc. Lond. A} \textbf{\bibinfo{volume}{242}},
  \bibinfo{pages}{300} (\bibinfo{year}{1957}).

\bibitem[{\citenamefont{Hanbury~Brown and Twiss}(1958)}]{HBT3}
\bibinfo{author}{\bibfnamefont{R.}~\bibnamefont{Hanbury~Brown}}
  \bibnamefont{and} \bibinfo{author}{\bibfnamefont{R.~Q.} \bibnamefont{Twiss}},
  \bibinfo{journal}{Proc. R. Soc. Lond. A} \textbf{\bibinfo{volume}{243}},
  \bibinfo{pages}{291} (\bibinfo{year}{1958}).

\bibitem[{\citenamefont{Alon et~al.}(2008)\citenamefont{Alon, Streltsov, and
  Cederbaum}}]{alon:08}
\bibinfo{author}{\bibfnamefont{O.~E.} \bibnamefont{Alon}},
  \bibinfo{author}{\bibfnamefont{A.~I.} \bibnamefont{Streltsov}},
  \bibnamefont{and} \bibinfo{author}{\bibfnamefont{L.~S.}
  \bibnamefont{Cederbaum}}, \bibinfo{journal}{Phys. Rev. A}
  \textbf{\bibinfo{volume}{77}}, \bibinfo{pages}{033613}
  (\bibinfo{year}{2008}).

\bibitem[{\citenamefont{Streltsov et~al.}(2013)\citenamefont{Streltsov,
  Sakmann, Lode, Alon, and Cederbaum}}]{Streltsov2010}
\bibinfo{author}{\bibfnamefont{A.~I.} \bibnamefont{Streltsov}},
  \bibinfo{author}{\bibfnamefont{K.}~\bibnamefont{Sakmann}},
  \bibinfo{author}{\bibfnamefont{A.~U.~J.} \bibnamefont{Lode}},
  \bibinfo{author}{\bibfnamefont{O.~E.} \bibnamefont{Alon}}, \bibnamefont{and}
  \bibinfo{author}{\bibfnamefont{L.~S.} \bibnamefont{Cederbaum}},
  \emph{\bibinfo{title}{The {M}ulticonfigurational time-dependent {H}artree for
  {B}osons package, version 2.3, {H}eidelberg}} (\bibinfo{year}{2013}),
  \urlprefix\url{http://mctdhb.org/}.

\end{thebibliography}

\end{document}